\documentclass[journal]{IEEEtran}
\ifCLASSINFOpdf
\else
\fi
\usepackage[ruled]{algorithm2e} 
\usepackage{caption}
\usepackage{subcaption}
\usepackage{amssymb}
\usepackage{gensymb}
\usepackage{graphicx}
\usepackage{pgfplots}
\usetikzlibrary{positioning}
\usepackage{tikz}
\usetikzlibrary{backgrounds}
\tikzset{>=latex}
\usetikzlibrary{arrows,shapes}
\usetikzlibrary{positioning,fit,calc}
\usepackage{epstopdf}
\usepackage{filecontents}

\SetAlFnt{\small}
\SetAlCapFnt{\small}
\SetAlCapNameFnt{\small}
\SetAlCapHSkip{0pt}
\IncMargin{-\parindent}
\usetikzlibrary{intersections}
\usepackage{multirow}	
\usepackage{xcolor,colortbl}
\usepackage{tabularx}
\usepackage{mathtools}
\usepackage[utf8]{inputenc}
\usepackage[english]{babel}
\usepackage{amsthm}

\makeatletter
\newcommand\resetstackedplots{
	\makeatletter
	\pgfplots@stacked@isfirstplottrue
	\makeatother
	\addplot [forget plot,draw=none] coordinates{(1,0) (2,0) (3,0) (4,0)};
}
\newcommand\resetthreestackedplots{
	\makeatletter
	\pgfplots@stacked@isfirstplottrue
	\makeatother
	\addplot [forget plot,draw=none] coordinates{(1,0) (2,0) (3,0)};
}
\makeatother
\pgfplotsset{compat=1.5}

\definecolor{green}{HTML}{8DB600}
\definecolor{red}{HTML}{E52B50}

\newtheorem{theorem}{Theorem}[section]
\newtheorem{lemma}[theorem]{Lemma}

\begin{document}
%
\title{Stark: Fast and Scalable Strassen's Matrix Multiplication using Apache Spark}
%
%
%

\author{Chandan~Misra,~Sourangshu~Bhattacharya,~and~Soumya~K.~Ghosh,~IIT Kharagpur
\thanks{C. Misra was with the Advanced Technology Development Centre, Indian Institute of Technology, Kharagpur,
West Bengal, 721302 INDIA e-mail: chandan.misra@iitkgp.ac.in}
\thanks{S. Bhattacharya and S. K. Ghosh are with Department of Computer Science and Engineering, Indian Institute of Technology Kharagpur
West Bengal, 721302 INDIA e-mail: sourangshu@cse.iitkgp.ernet.in and skg@iitkgp.ac.in}
\thanks{Manuscript received April 19, 2005; revised August 26, 2015.}}

%
%

\markboth{Journal of \LaTeX\ Class Files,~Vol.~14, No.~8, August~2015}%
{Shell \MakeLowercase{\textit{et al.}}: Bare Demo of IEEEtran.cls for IEEE Journals}
%



\maketitle

\begin{abstract}
This paper presents a new fast, highly scalable distributed matrix multiplication algorithm on Apache Spark, called \textit{Stark}, based on Strassen's matrix multiplication algorithm. Stark preserves Strassen's seven multiplications scheme in a distributed environment and thus achieves faster execution. It is based on two new ideas; it creates a recursion tree of computation where each level of such tree corresponds to division and combination of distributed matrix blocks in the form of \textit{Resilient Distributed Datasets} (RDDs); It processes each divide and combine step in parallel and memorize the sub-matrices by intelligently tagging matrix blocks in it. To the best of our knowledge, Stark is the first Strassen's implementation in Spark platform. We show experimentally that Stark has a strong scalability with increasing matrix size enabling us to multiply two ($16384\times 16384$) matrices with 28\% and 36\% less wall clock time than \textit{Marlin} and \textit{MLLib} respectively, state-of-the-art matrix multiplication approaches based on Spark.
\end{abstract}


%
\IEEEpeerreviewmaketitle

\section{Introduction}
%
%
%
%
\IEEEPARstart{G}{rowth} in the number of massive datasets from different sources like social media, weather sensors, mobile devices, etc. has led to applications of these datasets for various data-driven research and analytics in domains such as machine learning, climate science, social media analytics, etc. These applications require large-scale data processing with minimal effort and a system which scales as data grows without any failure. Many of these applications need matrix computations on massive datasets, leading to a requirement of large-scale distributed matrix computations.

Big data processing frameworks like Hadoop MapReduce \cite{hadoop} and Spark \cite{spark} have emerged as next-generation distributed programming platform for data-intensive complex analytics and developing distributed applications in fields like machine learning, climate science, and social media analytics. These reliable shared storage and analysis systems put the systems like RDBMS, grid computing and volunteer computing behind by its powerful batch processing, scalability and fault tolerance capabilities. Spark has gained its popularity for its in-memory data processing ability to run programs faster than Hadoop MapReduce. Its general purpose engine supports a wide range of applications including batch, interactive, iterative algorithms and streaming and it offers simple APIs and rich built-in libraries like MLLib \cite{meng2016mllib}, GraphX \cite{xin2013graphx} for data science tasks and data processing applications. Therefore, we can get substantial gain by implementing computing intensive algorithms which consume large dataset as input. In the present work, we focus on the problem of distributed multiplication of large and possibly distributed matrices using the Spark framework.

Many existing works have implemented distributed matrix multiplication on Big data frameworks. One of the early works was \textit{HAMA} \cite{seo2010hama}, which implemented distributed matrix multiplication on MapReduce. However, this scheme suffers from the shortcomings of Hadoop, i.e. communicating with HDFS for each map or reduce task. This drawback can be overcome by using the Spark framework, which supports distributed in-memory computation. The most widely used approach is the distributed matrix multiplication scheme used in its built-in machine learning library, called \textit{MLLib}. Another recent distributed matrix multiplication scheme is \textit{Marlin}\cite{gu2015efficient,gu2017improving}, which intelligently selects one of their three matrix multiplication algorithms according to the size of the input matrices. However, both these schemes used naive distributed block matrix multiplication approach. This approach requires $8$ block multiplications to calculate the product matrix when the input matrix is further divided in ($2\times 2$) blocks, which still requires $O(n^{3})$ running time. In the present work, we attempt to overcome this shortcoming by using Strassen's matrix multiplication algorithm, which was proposed by Volker Strassen in 1969 \cite{strassen1969gaussian}. Strassen's algorithm only needs $7$ block matrix multiplications for the ($2\times 2$) splitting of matrices, thus resulting in a time complexity of $O(n^{2.807})$. An interesting research question is whether this gain in complexity translates to gains in actual wall clock execution time on reasonably sized matrices when implemented using a Big data processing platform such as spark.

Strassen's algorithm, which is inherently recursive in nature, cannot be implemented efficiently in Hadoop MapReduce. This is because the MapReduce programming paradigm only supports stateless distributed operations so that fault tolerance can be ensured. Hence, for maintaining distributed states in Hadoop, one has to resort to disk-based data structures in HFDS or use external distributed key-value stores such as zookeeper, parameter server, etc. Spark is a natural choice since it can make recursive calls in the methods of driver program which can launch distributed in-memory jobs. The distributed state information can be stored as tags in the in-memory distributed data structure, thus supporting a more natural and efficient implementation for distributed recursion. Moreover, spark programs are part of the overall Hadoop ecosystem, hence interoperable with HDFS, Cassandra, HBase, Hive etc. Hence, our distributed matrix multiplication scheme can be used as part of larger data analytics workflows, where the input matrices are generated by some other Spark or MapReduce jobs and the product matrix from our technique can be consumed by some other jobs in the workflow.

There are several research challenges in developing the distributed version of Strassen's matrix multiplication algorithm, in the Map-Reduce framework:
\begin{itemize}
\item Strassen's algorithm is recursive in nature and thus not directly suitable for the Map-Reduce framework, which essentially assumes stateless functions for fault-tolerance. Hence, careful bookkeeping is needed for maintaining the state information in the global parameters, redundant distributed datasets (RDDs) in case of Spark.
\item The matrix is not easily partitionable i.e. each element in the product matrix depends on multiple elements in the input / intermediate matrices. Therefore, each partition cannot be processed independently which is one of the requirements for MapReduce programming model.
\item Even though Strassen's algorithm is theoretically faster, the tradeoffs between three key elements: \textit{computation}, \textit{communication} (I/O), and \textit{parallelism} (the number of actions happening parallelly) determine whether an actual speedup in wall clock time will be observed. In order to arrive at a  suitable tradeoff, careful theoretical analysis of different stage of execution for the distributed Strassen algorithm is needed in all three aspects.
\end{itemize}

In each Strassen's recursive call, the input matrices are divided into 7 sub-matrices and each such sub-matrix depends on the elements of components of other partitions. Also, it is necessary to keep track of the sub-matrices and matrix blocks in the intermediate map group and reduce phases, so that it can be further divided or merged and can thus get the final position in the product matrix. In this paper, we address the above challenges with a distributed tail recursion which is created by intelligently labeling the sub-matrices and matrix blocks for each recursive call. The tags are chosen in a way such that the division of the matrices can be done in parallel in a top-down fashion and also product sub-matrices can be arranged from the divisions in parallel as well in a bottom-up approach.
The distributed algorithm developed here was implemented using the Apache Spark framework algorithm. We performed extensive empirical evaluations to wall clock running time of our implementation vis-\`a-vis the state of the art implementations available on Spark (both \textit{MLLib} and \textit{Marlin}). We show that for a large range of practical matrix sizes, our implementation performs $40$\% -- $60$\% better than the nearest competitor.

Additionally, we report a comprehensive analysis of the \textit{computation complexity}, \textit{communication complexity}, and \textit{parallelization factor} for the implemented Strassen's algorithm as well as the baseline methods mentioned above. Similar analysis has also been reported by Gu et al. \cite{gu2015efficient}, though they do not explicitly report the analysis for different stages. This is critical in our case since the number of stages depends on the size of the matrix.

Since we are interested in the wall clock time, we do the above analysis for each stage of spark execution, which are executed serially. The wall clock time for each stage is governed by the dominant component (either computation or communication), and the parallelization factor which allows the total computation (or communication) for each stage to be divided into parallel executors. The total wall clock time is the sum of wall clock times of stages. We find that the careful theoretical computations, match the empirical observations of wall clock time when the implemented program are run in a distributed setting. Hence, this analysis helps us to pinpoint the source of the improvement in wall clock time. We find that the dominant component of running time in all the competing systems is the leaf node block multiplication that is executed in separate executors in parallel and our system outperformed them in the number of multiplications performed in leaf nodes. While \textit{MLLib} and \textit{Marlin} require $b^{3}$ ($b=partition size$) multiplications, our system needs only $b^{\log{7}}$ multiplications. For large enough matrices, matrix and block sizes differ significantly which provide us much steeper running time curve for our implementation.

We empirically demonstrate the effectiveness of our algorithm by comparing it to the best performing baselines over all possible partition sizes, for various matrix sizes. We find that our method takes 28\% and 36\% less wall clock time than \textit{Marlin} and \textit{MLLib} respectively.
In another experiment, we vary partition sizes for each matrix size. We find that running time follows a U-shaped pattern, thus suggesting an optimal block size for each matrix size, which is also intuitive. 

This experiment also shows that theoretically calculated running times and empirically observed running times match closely, hence further validating our theoretical calculations.
Finally, we report stage-wise breakup of both theoretical running times and empirically observed running times. This helps us to identify the most time-consuming stage, thus re-affirming our conclusion regarding the reason for improvement of running time with Strassen's algorithm over existing baselines.

\subsubsection{Organization of the article}
After presenting a detailed related work in section \ref{sec:related-work}, we introduce the Strassen's multiplication algorithm on a single node in section \ref{sec:Strassen-Multiplication-Preliminaries}. We introduce our algorithm \textit{Stark} from section \ref{block-data-structure} and provides a detailed description of the algorithm along with the data structure used. In section \ref{sec:Performance_Modeling}, we evaluate the performance analysis of our algorithm along with two other competing approaches --- \textit{MLLib} and \textit{Marlin} in order to show that \textit{Stark} has a better performance over others. This will also guide us to explain our experimental results provides in section \ref{sec:Experimental-Evaluation}. Section \ref{sec:conclusion} summarizes the results and discusses the future research direction.

\section{Related Work}
\label{sec:related-work}
An extensive literature exists on parallelizing naive matrix multiplication algorithms \cite{cannon1969cellular,berntsen1989communication}, \cite{choi1994pumma}, \cite{agarwal1995three}, \cite{van1997summa}, \cite{gu2015efficient}, \cite{choi1992scalapack}, \cite{dumitrescu1994fast}, \cite{mccoll1999memory}, \cite{hunold2008combining}, \cite{seo2010hama}, \cite{ballard2012graph}, \cite{solomonik2011communication}, \cite{demmel2013communication} and \cite{solomonik2012matrix}. Similarly Strassen's matrix multiplication algorithm has also been extensively studied for parallelization \cite{kumar1995tensor}, \cite{douglas1994gemmw}, \cite{grayson1996high}, \cite{luo1995scalable}, \cite{mccoll1999memory}, \cite{thottethodi1998tuning}, \cite{desprez2004impact}, \cite{ohtaki2004parallel}, \cite{song2006experiments}, \cite{lipshitz2012communication} and \cite{ballard2012communication}. As pointed out in \cite{gu2015efficient}, \cite{demmel2013communication} and \cite{lipshitz2012communication}, the literature on parallel and distributed matrix multiplication can be divided broadly into three categories: 1) Grid based approach, 2) BFS/DFS based approach and 3) Hadoop and Spark based approach. Here we briefly review them.

\subsection{Grid Based Approach}
The grid-based approaches are particularly well suited for the processor layout in a two or three-dimensional grid. In this layout, the communication occurs either in the same row or in the same column. Based on this processor layout, these approaches are again classified as 2D, 2.5D, and 3D. 2D and 3D cases treat processors as two and three-dimensional grid respectively. The most common known 2D algorithms are \cite{cannon1969cellular} and \cite{van1997summa}. In 3D approaches like \cite{agarwal1995three} and \cite{berntsen1989communication}, the communication cost is minimized using extra memory than 2D algorithms. It also reduces the bandwidth cost compared to 2D algorithms. 2.5D multiplication approach in \cite{solomonik2011improving}, \cite{solomonik2011communication} and \cite{mccoll1995bsp}, has been developed to interpolate between 2D and 3D approaches. It has a better bandwidth than 2D.

Strassen's matrix multiplication has also gone through a similar evolution and got new 2D and 2.5D approaches. Luo and Drake \cite{luo1995scalable} presented a scalable and parallel Strassen's matrix multiplication algorithm. They have provided two approaches to multiply two matrices in parallel. The first approach is to use classical parallel matrix multiply for the parallelization and Strassen's multiply method locally --- called the 2D-Strassen. In the second approach, they reversed the order i.e. they parallelize Strassen's at the higher levels and use standard parallel matrix multiplication at lower levels --- called the Strassen-2D. They analyzed the communication costs for these two approaches. Grayson et. al. \cite{grayson1996high} improved on the second approach and concluded that the second one is the best approach under their new circumstances. Then comes the 2.5D version of the 2D-Strassen and Strassen-2D algorithms. In \cite{solomonik2011improving}, they got better communication efficiency than its 2D counterparts but still lacks communication optimality. Grid-based algorithms are very efficient in a grid and torus-based topologies but may not perform well in other more general topologies \cite{demmel2013communication}, which is the main focus of generic Big Data computing platforms.

\subsection{BFS/DFS Based Approach}
The failure of the above-mentioned approaches to achieve communication optimality, BFS/DFS approach is developed \cite{ballard2012communication} for Strassen's algorithm. BFS/DFS approach treats processor layout as a hierarchy rather than two or three-dimensional grid and based on a sequential recursive algorithm. Among Strassen based other parallel algorithms, (Communication-Optimal Parallel Strassen's) \textit{CAPS} \cite{lipshitz2012communication} provides the minimized communication costs and runs faster in practice. Ballard et. al. presented the communication costs for Strassen in \cite{ballard2012graph} and \cite{ballard2012communication} and also provides the communication lower bound for square as well as for rectangular matrices in \cite{demmel2013communication}. \textit{CAPS} matches the lower bound and provides communication optimality. 

\textit{CAPS} traverses the Strassen recursion tree in parallel in two ways. In the \textit{unlimited memory} (UM) scheme, it takes $k$ BFS steps and then performs local matrix multiplication. The \textit{Limited Memory} approach takes $l$ DFS steps and then $k$ BFS steps. The memory footprint can be minimized by minimizing $l$. They also showed that second approach can be tuned to get more complicated interleaving approach but does not attain optimality more than a constant factor. Though our implementation follows a similar kind of recursion tree as \textit{CAPS}, it is worth evaluating the algorithm in a scalable framework where data is distributed.

\subsection{Hadoop and Spark Based Approach}
There are several implementations of distributed matrix multiplication on using Hadoop MapReduce and Spark. John Norstad in \cite{norstad} presented four strategies to implement data parallel matrix multiplication using block matrix data structure. However, all of them are based on the classical parallel approach which requires eight multiplications.

There are other distributed framework that provides massive matrix computation like \textit{HAMA} \cite{seo2010hama} and \textit{MadLINQ} \cite{qian2012madlinq}. matrix multiplication in \textit{HAMA} is carried out using two approaches --- \textit{iterative} and \textit{Block}. In the iterative approach, each map task receives a row index of the right matrix as a key and the column vector of the row as a value. Then it multiplies all columns of $i^{th}$ row of the left matrix with the received column vector. Finally, a reduce task collects the $i^{th}$ product into the result matrix. \textit{Block} approach reduces required data movement over the network by building a collection table and placing candidate block matrix in each row. However, the iterative approach is not suitable in Hadoop for massive communication cost. Though \textit{Block} approach incurs low communication cost, does not provide faster execution as it uses classical parallel matrix multiplication approach. \textit{MadLINQ}, built on top of Microsoft's \textit{LINQ} framework and \textit{Dryad} \cite{isard2007dryad}, is an example of cloud-based linear algebra platform. However, it suffers from the same kind of drawback as \textit{HAMA}.

Rong Gu et al. in \cite{gu2015efficient} developed an efficient distributed computation library, called \textit{Marlin}, on top of Apache Spark. They proposed three different matrix multiplication algorithm based on the size of input matrices. They have shown that Marlin is faster than R and distributed algorithms based on MapReduce. When the matrix sizes are square, they have used a hybrid of the naive block matrix multiplication scheme. Though they have minimized the shuffle in join step, underlying they incur 8 multiplications compared to 7 multiplications on Stark, which makes Stark faster than Marlin. \textit{MLLib} block matrix multiplication does the same thing, but a little bit different way. The algorithm first lists all the partitions for each block that are needed in the same place and then shuffles, which reduce the communication cost.

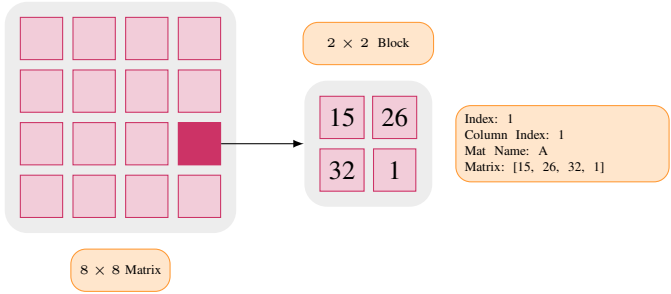
\begin{figure}[!ht]		
	\centering
		\begin{tikzpicture}[squarednode/.style={rectangle, draw=purple!80, fill=purple!20, thin, minimum size=1.6em},squarednodeFill/.style={rectangle, draw=purple!80, fill=purple!80, thin, minimum size=1.6em}, squarednodeDetails/.style={rectangle, draw=orange!80, fill=orange!20, thin, minimum size=1.6em}, squarednodeNoFill/.style={rectangle, draw=white!80, fill=white!10, thin, minimum size=1.6em}, background/.style={rectangle,fill=gray!14,inner sep=0.2cm,rounded corners=3mm}, node distance=0.7cm,->]
		\node[squarednode](00){};
		\node[squarednode](10)[below of = 00]{};
		\node[squarednode](20)[below of = 10]{};
		\node[squarednode](30)[below of = 20]{};
		
		\node[squarednode](01)[right of = 00]{};
		\node[squarednode](11)[right of = 10]{};
		\node[squarednode](21)[right of = 20]{};
		\node[squarednode](31)[right of = 30]{};
		
		\node[squarednode](02)[right of = 01]{};
		\node[squarednode](12)[right of = 11]{};
		\node[squarednode](22)[right of = 21]{};
		\node[squarednode](32)[right of = 31]{};
		
		\node[squarednode](03)[right of = 02]{};
		\node[squarednode](13)[right of = 12]{};
		\node[squarednodeFill](23)[right of = 22]{};
		\node[squarednode](33)[right of = 32]{};
		
		\begin{pgfonlayer}{background}
		\node [background,fit= (00)(01)(02)(03)(10)(11)(12)(13)(20)(21)(22)(23)(30)(31)(32)(33)] (LargeBox) {};
		\end{pgfonlayer}
		
		\node[squarednodeDetails](LargeMatrixTitle)[below = 0.2cm of LargeBox, rounded corners=0.2cm]{\tiny $8\times 8$ Matrix};
		
		\node[squarednode](small00)[right = 1.1cm of LargeBox]{15};
		\node[squarednode](small01)[right of = small00]{26};
		\node[squarednode](small10)[below of = small00]{32};
		\node[squarednode](small11)[right of = small10]{1};
		
		\begin{pgfonlayer}{background}
		\node [background,fit= (small00)(small01)(small10)(small11)] (SmallBox) {};
		\end{pgfonlayer}
		
		\node[squarednodeDetails](SmallMatrixTitle)[above = 0.2cm of SmallBox, text width=1.5cm, align = center, rounded corners=0.2cm]{\baselineskip=5pt\tiny $2\times 2$ Block\par};
		
		\draw (23.east) -- (SmallBox.west);
		
		\node[squarednodeDetails](details)[right = 0.3cm of SmallBox, text width=2.6cm, rounded corners=0.2cm]{\baselineskip=5pt\tiny Index: 1 \\ Column Index: 1 \\ Mat Name: A \\ Matrix: [15, 26, 32, 1]\par};
		
		\end{tikzpicture}		
		\caption{Matrix Block Data Structure. Each block consists of a submatrix of dimension $>= 2$. Figure above shows a matrix of size $(8 \times 8)$ having $(4 \times 4)$ matrix blocks each of which is a $(2 \times 2)$ sub-matrix.}
		\label{fig:matrix-block}
	\end{figure}

\section{Distributed Strassen's on Spark}
\label{sec:Distributed-Strassen-Implementation}
In this section, we discuss the implementation of \textit{Stark} on Spark framework. First, we describe the original Strassen's algorithm for serial matrix multiplication in section \ref{sec:Strassen-Multiplication-Preliminaries}. Next, we describe the \textit{block} data structure, which is central to our distributed matrix multiplication algorithm, since it encapsulates both the contents of a block as well as \textit{tags} necessary for the distributed recursive algorithm in section \ref{block-data-structure}. Finally, section \ref{sec:implementation-details} describes the distributed matrix multiplication algorithm, and its implementation strategy using RDD (resilient distributed datasets construct in Spark \cite{zaharia2010spark}) of blocks.

\subsection{Single Node Strassen's Preliminaries}
\label{sec:Strassen-Multiplication-Preliminaries}
Strassen's matrix multiplication can multiply two $(n \times n)$ matrix using $7$ multiplications and $18$ additions of matrix blocks of size $\frac{n}{2}\times\frac{n}{2}$, thus providing much faster execution compared to $8$ multiplications and $4$ additions of the naive algorithm. Algorithm \ref{strassen-algo} lists the scheme.

Note that $n$ should be $2^p$ for some integer $p$. However, the scheme can also be applied to rectangular matrices or matrices of general sizes by partitioning them appropriately as given in \cite{luo1995scalable} to be used in practical applications. In this paper, we focus on matrices of size $2^p$ for mathematical brevity.

\begin{algorithm}[!ht]
	\SetAlgoLined
	\SetKwFunction{proc}{Strassen's}
	\SetKwProg{myproc}{Procedure}{}{}
	\myproc{\proc{A, B, threshold}}{
		A = input matrix of size $n \times n$\;
		B = input matrix of size $n \times n$\;
		C = input matrix of size $n \times n$\;
		\eIf{n=thresold}{
			Multiply A and B using naive approach\;
		}{
			Compute $A_{11}, B_{11}, ..., A_{22}, B_{22}$ by computing $n=\frac{n}{2}$\;
			$M_{1} =$ \textsc{Strassen's$((A_{11}+A_{22})$,$(B_{11}+B_{22}))$}\;
			$M_{2} =$ \textsc{Strassen's$((A_{21}+A_{22})$,$B_{11})$}\;
			$M_{3} =$ \textsc{Strassen's$(A_{11}$,$(B_{12}-B_{22})$}\;
			$M_{4} =$ \textsc{Strassen's$(A_{22}$,$(B_{21}-B_{11}))$}\;
			$M_{5} =$ \textsc{Strassen's$((A_{11}+A_{12})$,$B_{22})$}\;
			$M_{6} =$ \textsc{Strassen's$((A_{21}-A_{11})$,$(B_{11}+B_{12}))$}\;
			$M_{7} =$ \textsc{Strassen's$((A_{12}-A_{22})$,$(B_{21}+B_{22}))$}\;
			
			$C_{11} = (M_{1}+M_{4}-M_{5}+M_{7})$\;
			$C_{12} = (M_{3}+M_{5})$\;
			$C_{21} = (M_{2}+M_{4})$\;
			$C_{22} = (M_{1}-M_{2}-M_{3}+M_{6})$\;
		}
		\KwRet C\;}    
	\caption{Strassen's Matrix Multiplication}\label{strassen-algo}
\end{algorithm}


\subsection{Block Data Structure}
\label{block-data-structure}
The main data structure used is a matrix, which is represented as an RDD of \textit{blocks}. Blocks store information necessary for (1) representing the matrix i.e. storing all the entries and (2) bookkeeping information needed for running the algorithm.

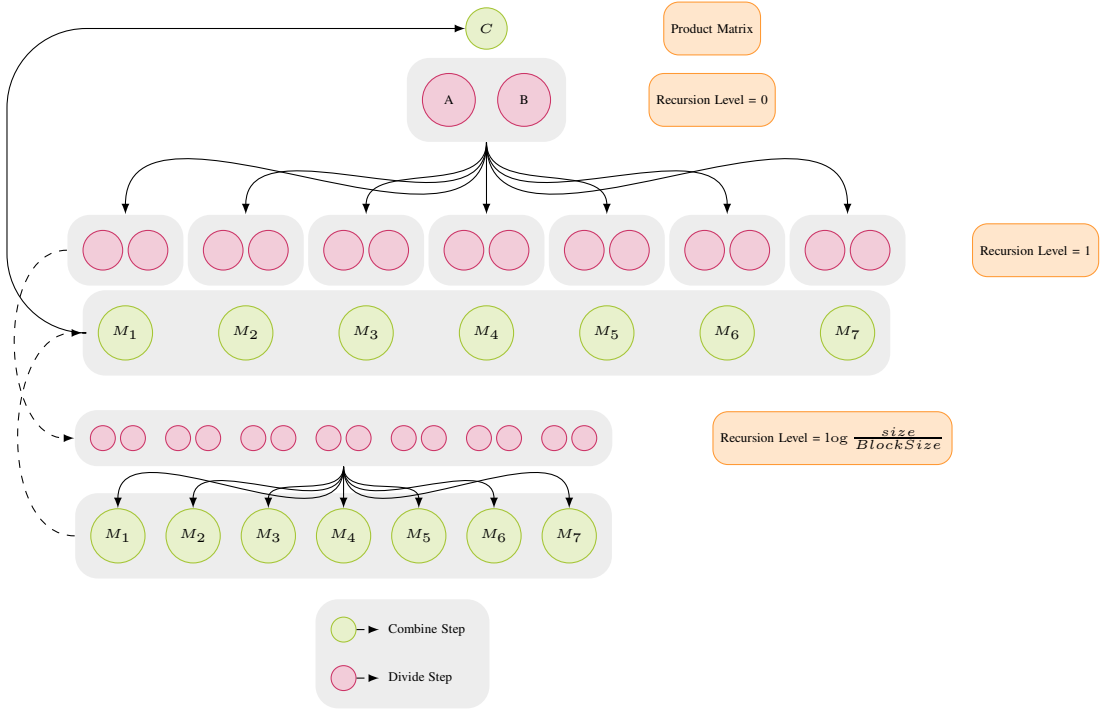
\begin{figure*}[!ht]		
	\centering
	\begin{tikzpicture}[circlenodeBig/.style={circle, draw=purple!80, fill=purple!20, thin, minimum size=2em},rectangleBox/.style={rectangle, draw=orange!80, fill=orange!20, thin, minimum size=2em, rounded corners=2mm, inner sep=0.1cm},circlenode/.style={circle, draw=purple!80, fill=purple!20, thin, minimum size=1.5em},circlenodesmall/.style={circle, draw=purple!80, fill=purple!20, thin, minimum size=0.1em}, rectanglenodeEmpty/.style={rectangle}, circlenodesmallGreen/.style={circle, draw=green!80, fill=green!20, thin, minimum size=0.1em},background/.style={rectangle,fill=gray!14,inner sep=0.2cm,rounded corners=3mm},backgroundEx/.style={rectangle,fill=gray!14,inner sep=0.4cm,rounded corners=3mm},rectangleconnector/.style={draw,to path={(\tikztostart) -- ++(#1,0pt) \tikztonodes |- (\tikztotarget) },pos=0.5},->]
	
	\node[circlenodeBig](A){\tiny A};
	\node[circlenodeBig](B)[right of = A]{\tiny B};
	
	\begin{pgfonlayer}{background}
	\node [background,fit= (A)(B)] (level1) {};
	\end{pgfonlayer}				
	
	\node[circlenode](M1A)[below of = A, xshift = -4.6cm, yshift = -1cm]{};
	\node[circlenode](M1B)[below of = B, xshift = -5cm, yshift = -1cm]{};
	\node[circlenode](M2A)[below of = A, xshift = -3cm, yshift = -1cm]{};
	\node[circlenode](M2B)[below of = B, xshift = -3.4cm, yshift = -1cm]{};
	\node[circlenode](M3A)[below of = A, xshift = -1.4cm, yshift = -1cm]{};
	\node[circlenode](M3B)[below of = B, xshift = -1.8cm, yshift = -1cm]{};
	\node[circlenode](M4A)[below of = A, xshift = 0.2cm, yshift = -1cm]{};
	\node[circlenode](M4B)[below of = B, xshift = -0.2cm, yshift = -1cm]{};
	\node[circlenode](M5A)[below of = A, xshift = 1.8cm, yshift = -1cm]{};
	\node[circlenode](M5B)[below of = B, xshift = 1.4cm, yshift = -1cm]{};
	\node[circlenode](M6A)[below of = A, xshift = 3.4cm, yshift = -1cm]{};
	\node[circlenode](M6B)[below of = B, xshift = 3cm, yshift = -1cm]{};
	\node[circlenode](M7A)[below of = A, xshift = 5cm, yshift = -1cm]{};
	\node[circlenode](M7B)[below of = B, xshift = 4.6cm, yshift = -1cm]{};
	
	\begin{pgfonlayer}{background}
	\node [background,fit= (M1A)(M1B)] (level21) {};
	\node [background,fit= (M2A)(M2B)] (level22) {};
	\node [background,fit= (M3A)(M3B)] (level23) {};
	\node [background,fit= (M4A)(M4B)] (level24) {};
	\node [background,fit= (M5A)(M5B)] (level25) {};
	\node [background,fit= (M6A)(M6B)] (level26) {};
	\node [background,fit= (M7A)(M7B)] (level27) {};
	\end{pgfonlayer}
	
	\node[circlenodesmallGreen](M1)[below of = M1A, xshift=0.3cm, yshift=-0.1cm]{\tiny $M_{1}$};
	\node[circlenodesmallGreen](M2)[below of = M2A, xshift=0.3cm, yshift=-0.1cm]{\tiny $M_{2}$};
	\node[circlenodesmallGreen](M3)[below of = M3A, xshift=0.3cm, yshift=-0.1cm]{\tiny $M_{3}$};
	\node[circlenodesmallGreen](M4)[below of = M4A, xshift=0.3cm, yshift=-0.1cm]{\tiny $M_{4}$};
	\node[circlenodesmallGreen](M5)[below of = M5A, xshift=0.3cm, yshift=-0.1cm]{\tiny $M_{5}$};
	\node[circlenodesmallGreen](M6)[below of = M6A, xshift=0.3cm, yshift=-0.1cm]{\tiny $M_{6}$};
	\node[circlenodesmallGreen](M7)[below of = M7A, xshift=0.3cm, yshift=-0.1cm]{\tiny $M_{7}$};
	
	\node[circlenodesmall](LeafM1A)[below of = M1A, xshift=0cm, yshift=-1.5cm]{};
	\node[circlenodesmall](LeafM1B)[right of = LeafM1A, xshift=-0.6cm]{};
	\node[circlenodesmall](LeafM2A)[right of = LeafM1B, xshift=-0.4cm]{};
	\node[circlenodesmall](LeafM2B)[right of = LeafM2A, xshift=-0.6cm]{};
	\node[circlenodesmall](LeafM3A)[right of = LeafM2B, xshift=-0.4cm]{};
	\node[circlenodesmall](LeafM3B)[right of = LeafM3A, xshift=-0.6cm]{};
	\node[circlenodesmall](LeafM4A)[right of = LeafM3B, xshift=-0.4cm]{};
	\node[circlenodesmall](LeafM4B)[right of = LeafM4A, xshift=-0.6cm]{};
	\node[circlenodesmall](LeafM5A)[right of = LeafM4B, xshift=-0.4cm]{};
	\node[circlenodesmall](LeafM5B)[right of = LeafM5A, xshift=-0.6cm]{};
	\node[circlenodesmall](LeafM6A)[right of = LeafM5B, xshift=-0.4cm]{};
	\node[circlenodesmall](LeafM6B)[right of = LeafM6A, xshift=-0.6cm]{};
	\node[circlenodesmall](LeafM7A)[right of = LeafM6B, xshift=-0.4cm]{};
	\node[circlenodesmall](LeafM7B)[right of = LeafM7A, xshift=-0.6cm]{};		
	
	\begin{pgfonlayer}{background}
	\node [background,fit= (LeafM1A)(LeafM1B)(LeafM2A)(LeafM2B)(LeafM3A)(LeafM3B)(LeafM4A)(LeafM4B)(LeafM5A)(LeafM5B)(LeafM6A)(LeafM6B)(LeafM7A)(LeafM7B)] (levelLeaf) {};
	\end{pgfonlayer}					
	
	\node[circlenodesmallGreen](LeafM1)[below of = LeafM1A, xshift=0.2cm, yshift=-0.3cm]{\tiny $M_{1}$};
	\node[circlenodesmallGreen](LeafM2)[right of = LeafM1]{\tiny $M_{2}$};
	\node[circlenodesmallGreen](LeafM3)[right of = LeafM2]{\tiny $M_{3}$};
	\node[circlenodesmallGreen](LeafM4)[right of = LeafM3]{\tiny $M_{4}$};
	\node[circlenodesmallGreen](LeafM5)[right of = LeafM4]{\tiny $M_{5}$};
	\node[circlenodesmallGreen](LeafM6)[right of = LeafM5]{\tiny $M_{6}$};
	\node[circlenodesmallGreen](LeafM7)[right of = LeafM6]{\tiny $M_{7}$};					
	
	\begin{pgfonlayer}{background}
	\node [background,fit= (LeafM1)(LeafM1)(LeafM2)(LeafM2)(LeafM3)(LeafM3)(LeafM4)(LeafM4)(LeafM5)(LeafM5)(LeafM6)(LeafM6)(LeafM7)(LeafM7)] (SmallBox) {};
	\end{pgfonlayer}								
	
	\begin{pgfonlayer}{background}
	\node [background,fit= (M1)(M2)(M3)(M4)(M5)(M6)(M7)] (levelM1) {};
	\end{pgfonlayer}				
	
	\node[circlenodesmallGreen](C)[above of = A, xshift=0.5cm, yshift=-0.05cm]{\tiny $C$};
	
	\node[rectangleBox](productMatrix)[right of = C, xshift=2cm]{\tiny Product Matrix};
	\node[rectangleBox](recursionLevel0)[right of = B, xshift=1.5cm]{\tiny Recursion Level = 0};
	\node[rectangleBox](recursionLevel1)[right of = level27, xshift=1.5cm]{\tiny Recursion Level = 1};
	\node[rectangleBox](recursionLevel2)[right of = levelLeaf, xshift=5.5cm]{\tiny Recursion Level = $\log{\frac{size}{Block Size}}$};
	
	\node[circlenodesmallGreen](infoGreen1)[below = 0.5cm of SmallBox]{};
	\node[rectanglenodeEmpty](infoGreen2)[right = 0.3cm of infoGreen1]{\tiny Combine Step};
	\node[circlenodesmall](infoPurple1)[below = 0.3cm of infoGreen1]{};
	\node[rectanglenodeEmpty](infoPurple2)[right = 0.3cm of infoPurple1]{\tiny Divide Step};
	
	\draw [dashed,->](infoGreen1.east) to (infoGreen2.west);
	\draw [dashed,->](infoPurple1.east) to (infoPurple2.west);
	
	\begin{pgfonlayer}{background}
	\node [background,fit= (infoGreen1)(infoGreen2)(infoPurple1)(infoPurple2)] (info) {};
	\end{pgfonlayer}	
	
	\draw (level1.south) to [out=270, in=90](level21.north);
	\draw (level1.south) to [out=270, in=90](level22.north);
	\draw (level1.south) to [out=270, in=90](level23.north);
	\draw (level1.south) to [out=270, in=90](level24.north);
	\draw (level1.south) to [out=270, in=90](level25.north);
	\draw (level1.south) to [out=270, in=90](level26.north);
	\draw (level1.south) to [out=270, in=90](level27.north);
	
	\draw[dashed,->] (level21.west) to [out=180, in=180](levelLeaf.west);
	
	\draw (levelLeaf.south) to [out=270, in=90](LeafM1.north);
	\draw (levelLeaf.south) to [out=270, in=90](LeafM2.north);
	\draw (levelLeaf.south) to [out=270, in=90](LeafM3.north);
	\draw (levelLeaf.south) to [out=270, in=90](LeafM4.north);
	\draw (levelLeaf.south) to [out=270, in=90](LeafM5.north);
	\draw (levelLeaf.south) to [out=270, in=90](LeafM6.north);	
	\draw (levelLeaf.south) to [out=270, in=90](LeafM7.north);	
	
	\draw[dashed,->] (SmallBox.west) to [out=180, in=180] (levelM1.west);
	\draw[rectangleconnector=-1cm, rounded corners=30pt] (levelM1.west) to (C.west);
	\end{tikzpicture}		
	\caption{The implementation flow of Stark. Red circles with solid lines denote the division and replication of sub-matrices. Green circles with dashed lines denote the sub-matrices resulted from the combination phase of the recursion algorithm. Each recursion level (from $0$ to $\log{(size/BlockSize)}$) is executed in parallel.}
	\label{fig:distributedStrassen}
\end{figure*}

Conceptually, each matrix of dimension $n$, is partitioned into $b$ partitions, giving $\frac{n}{b}$ \textit{block rows} and $\frac{n}{b}$ \textit{block columns}. Here, \textit{block rows} and \textit{block columns} are defined by the number of rows and columns of blocks.

Each matrix of size $n$ is divided into four equal square sub-matrices of dimension $\frac{n}{2}$, until it reaches \textit{block} dimension of $\frac{n}{b}$. These sub-matrices are stored in data structure called \textit{blocks}, which is central to our algorithm. Note that, these \textit{blocks} are of fixed size, and can be stored and multiplied on a single node. Each block contains four fields (depicted in Fig. \ref{fig:matrix-block}):

\begin{enumerate}
    \item \texttt{row-index}: Stores current row index of the sub-matrix, in multiples of $n$. Note that, as the larger matrix is split during execution of the algorithm, these indices can change to keep track of the current position of sub-matrix.
    \item \texttt{column-index}: Similar to above, stores the current column index of the sub-matrix.
    \item \texttt{mat-name}: Stores a tag which is used as a key for grouping the blocks at each stage of the algorithm, so that blocks which need to be operated on are in the same group. It consists of a comma-separated string which denotes two components:
    \begin{enumerate}
        \item The matrix tag: stores the matrix label, for example, $A$ or $B$ or one of the eight sub-matrices $A_{11}$, $A_{12}$, $A_{21}$, $A_{22}$, $B_{11}$, $B_{12}$, $B_{21}$ and $B_{22}$ or $M$.
        \item M-Index: Each sub-matrix is broken down into $7$ sub-matrices. Therefore, this index helps to signify one of these $7$ sub-matrices.
    \end{enumerate}
    \item \texttt{matrix}: 2D array storing the matrix.
\end{enumerate}

\subsection{Implementation Details}
\label{sec:implementation-details}
The core multiplication algorithm (described in Algorithm \ref{alg:Distributed-Strassen's}) takes two matrices (say $A$ and $B$) represented as \textit{RDD} of blocks, as input as shown in Fig. \ref{fig:distributedStrassen}. The computation performed by the algorithm can be divided into 3 phases:

\begin{itemize}
    \item Recursively splitting each input matrix into $4$ equal sub-matrices and replicate the sub-matrices so as to facilitate the computation of intermediate matrices ($M_{1}$ to $M_{7}$).
    \item Multiply blocks serially to form blocks $M_{1}$ to $M_{7}$.
    \item Combine the sub-matrices to form matrices $C_{1}$ to $C_{4}$ of size $2^{n}$ from $2^{n-1}$.
\end{itemize}

Each step mentioned above run in parallel inside the cluster. Each step is described in the following sections.

\begin{algorithm}
\SetAlgoLined
\SetKwFunction{procone}{DistStrass}
\SetKwFunction{proctwo}{MulBlockMat}
\SetKwFunction{procthree}{DivNRep}
\SetKwFunction{procfour}{Combine}
\SetKwProg{myproc}{Procedure}{}{}
\myproc{\procone{$RDD<Block> A$, $RDD<Block> B$, int $n$}}{
    \KwResult{RDD of blocks of the product matrix C}
    $size =$ Size of matrix $A$ or $B$\;
    $blockSize =$ Size of a single matrix block\;
    $n =$ $\frac{size}{blockSize}$\;
    \eIf{n=1}{
        \tcc{Boundery Condition: RDD A and B contain blocks with a pair of blocks (candidates for multiplication) having same matname property}
        \proctwo{A,B}\;
    }{
        $n = \frac{n}{2}$\;
        \tcc{Divide Matrix A and B into 4 sub-matrices each ($A_{11}$ to $B_{22}$). Replicate and add or subtract the sub-matrices so that they can form 7 sub-matrices ($M_{1}$ to $M_{7}$)}
        D = \procthree{A,B}\;
        \tcc{Recursively call \procone{} to multiply two sub-matrices of block size $\frac{n}{2}$}
        R = \procone{A,B,n}\;
        \tcc{Combine seven submatrices (single RDD of blocks ($M_{1}$ to $M_{7}$)) of size $\frac{n}{2}$ into single matrix (RDD of blocks ($C$))}
        C = \procfour{R}\;
    }
    \KwRet C\;}
\caption{Distributed Strassen's Matrix Multiplication}
\label{alg:Distributed-Strassen's}
\end{algorithm}

\subsubsection{Divide and Replication Phase}
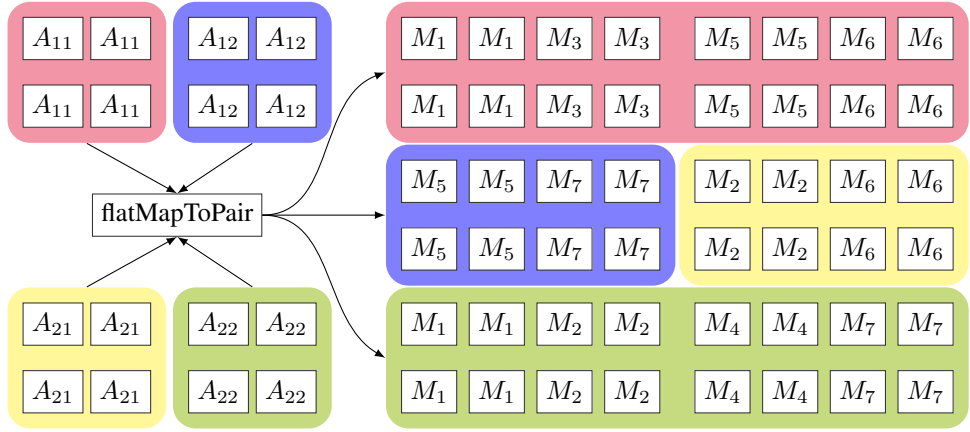
\begin{figure*}[!ht]
	\centering
		\begin{tikzpicture}[squarednodePurple/.style={rectangle, draw=purple!80, fill=purple!20, thin, minimum size=1.6em}, 
		squarednodeGreen/.style={rectangle, draw=green!80, fill=green!20, thin, minimum size=1.6em},
		squarednodeBlue/.style={rectangle, draw=blue!80, fill=blue!50, thin, minimum size=1.6em},
		squarednodeOrange/.style={rectangle, draw=yellow!80, fill=orange!20, thin, minimum size=1.6em},
		squarednode/.style={rectangle, draw=black!80, fill=white!20, thin, minimum size=1.6em},
		backgroundPurple/.style={rectangle,fill=red!50,inner sep=0.2cm,rounded corners=3mm},
		backgroundBlue/.style={rectangle,fill=blue!50,inner sep=0.2cm,rounded corners=3mm},
		backgroundOrange/.style={rectangle,fill=yellow!50,inner sep=0.2cm,rounded corners=3mm},
		backgroundGreen/.style={rectangle,fill=green!50,inner sep=0.2cm,rounded corners=3mm},
		node distance=0.9cm,->]
		
			\node[squarednode](00){$A_{11}$};
			\node[squarednode](10)[below of = 00]{$A_{11}$};
			\node[squarednode](20)[below of = 10, yshift = -2cm]{$A_{21}$};
			\node[squarednode](30)[below of = 20]{$A_{21}$};
			
			\node[squarednode](01)[right of = 00]{$A_{11}$};
			\node[squarednode](11)[right of = 10]{$A_{11}$};
			\node[squarednode](21)[right of = 20]{$A_{21}$};
			\node[squarednode](31)[right of = 30]{$A_{21}$};
			
			\node[squarednode](02)[right of = 01, xshift = 0.4cm]{$A_{12}$};
			\node[squarednode](12)[right of = 11, xshift = 0.4cm]{$A_{12}$};
			\node[squarednode](22)[below of = 12, yshift=-2cm]{$A_{22}$};
			\node[squarednode](32)[below of = 22] {$A_{22}$};
			
			\node[squarednode](03)[right of = 02]{$A_{12}$};
			\node[squarednode](13)[right of = 12]{$A_{12}$};
			\node[squarednode](23)[below of = 13, yshift=-2cm]{$A_{22}$};
			\node[squarednode](33)[below of = 23]{$A_{22}$};
			
			\begin{pgfonlayer}{background}
				\node [backgroundPurple,fit= (00)(01)(10)(11)] (BackgroundPurple) {};
				\node [backgroundBlue,fit= (02)(03)(12)(13)] (BackgroundBlue) {};
				\node [backgroundOrange,fit= (20)(21)(30)(31)] (BackgroundOrange) {};
				\node [backgroundGreen,fit= (22)(23)(32)(33)] (BackgroundGreen) {};
			\end{pgfonlayer}
			
			\node[squarednode](fl00)[right of = 03, xshift = 1cm]{$M_{1}$};
			\node[squarednode](fl01)[right of = fl00]{$M_{1}$};
			\node[squarednode](fl02)[right of = fl01]{$M_{3}$};
			\node[squarednode](fl03)[right of = fl02]{$M_{3}$};
			\node[squarednode](fl04)[right of = fl03, xshift = 0.3cm]{$M_{5}$};
			\node[squarednode](fl05)[right of = fl04]{$M_{5}$};
			\node[squarednode](fl06)[right of = fl05]{$M_{6}$};
			\node[squarednode](fl07)[right of = fl06]{$M_{6}$};
			
			\node[squarednode](fl10)[right of = 13, xshift = 1cm]{$M_{1}$};
			\node[squarednode](fl11)[right of = fl10]{$M_{1}$};
			\node[squarednode](fl12)[right of = fl11]{$M_{3}$};
			\node[squarednode](fl13)[right of = fl12]{$M_{3}$};
			\node[squarednode](fl14)[right of = fl13, xshift = 0.3cm]{$M_{5}$};
			\node[squarednode](fl15)[right of = fl14]{$M_{5}$};
			\node[squarednode](fl16)[right of = fl15]{$M_{6}$};
			\node[squarednode](fl17)[right of = fl16]{$M_{6}$};
			
			\begin{pgfonlayer}{background}
				\node [backgroundPurple,fit= (fl00)(fl01)(fl02)(fl03)(fl04)(fl05)(fl06)(fl07)(fl10)(fl11)(fl12)(fl13)(fl14)(fl15)(fl16)(fl17)] (BackgroundPurpleLarge) {};
			\end{pgfonlayer}
			
			\node[squarednode](fl20)[below of = fl10, yshift = -0.1cm]{$M_{5}$};
			\node[squarednode](fl21)[right of = fl20]{$M_{5}$};
			\node[squarednode](fl22)[right of = fl21]{$M_{7}$};
			\node[squarednode](fl23)[right of = fl22]{$M_{7}$};
			\node[squarednode](fl24)[right of = fl23, xshift = 0.3cm]{$M_{2}$};
			\node[squarednode](fl25)[right of = fl24]{$M_{2}$};
			\node[squarednode](fl26)[right of = fl25]{$M_{6}$};
			\node[squarednode](fl27)[right of = fl26]{$M_{6}$};
			
			\node[squarednode](fl30)[below of = fl20]{$M_{5}$};
			\node[squarednode](fl31)[right of = fl30]{$M_{5}$};
			\node[squarednode](fl32)[right of = fl31]{$M_{7}$};
			\node[squarednode](fl33)[right of = fl32]{$M_{7}$};
			\node[squarednode](fl34)[right of = fl33, xshift = 0.3cm]{$M_{2}$};
			\node[squarednode](fl35)[right of = fl34]{$M_{2}$};
			\node[squarednode](fl36)[right of = fl35]{$M_{6}$};
			\node[squarednode](fl37)[right of = fl36]{$M_{6}$};
			
			\begin{pgfonlayer}{background}
				\node [backgroundBlue,fit= (fl20)(fl21)(fl22)(fl23)(fl30)(fl31)(fl32)(fl33)] (BackgroundBlueLarge) {};
				\node [backgroundOrange,fit= (fl24)(fl25)(fl26)(fl27)(fl34)(fl35)(fl36)(fl37)] (BackgroundOrangeLarge) {};
			\end{pgfonlayer}
			
			\node[squarednode](fl40)[below of = fl30, yshift = -0.1cm]{$M_{1}$};
			\node[squarednode](fl41)[right of = fl40]{$M_{1}$};
			\node[squarednode](fl42)[right of = fl41]{$M_{2}$};
			\node[squarednode](fl43)[right of = fl42]{$M_{2}$};
			\node[squarednode](fl44)[right of = fl43, xshift = 0.3cm]{$M_{4}$};
			\node[squarednode](fl45)[right of = fl44]{$M_{4}$};
			\node[squarednode](fl46)[right of = fl45]{$M_{7}$};
			\node[squarednode](fl47)[right of = fl46]{$M_{7}$};
			
			\node[squarednode](fl50)[below of = fl40]{$M_{1}$};
			\node[squarednode](fl51)[right of = fl50]{$M_{1}$};
			\node[squarednode](fl52)[right of = fl51]{$M_{2}$};
			\node[squarednode](fl53)[right of = fl52]{$M_{2}$};
			\node[squarednode](fl54)[right of = fl53, xshift = 0.3cm]{$M_{4}$};
			\node[squarednode](fl55)[right of = fl54]{$M_{4}$};
			\node[squarednode](fl56)[right of = fl55]{$M_{7}$};
			\node[squarednode](fl57)[right of = fl56]{$M_{7}$};
			
			\begin{pgfonlayer}{background}
				\node [backgroundGreen,fit= (fl40)(fl41)(fl42)(fl43)(fl44)(fl45)(fl46)(fl47)(fl50)(fl51)(fl52)(fl53)(fl54)(fl55)(fl56)(fl57)] (BackgroundGreenLarge) {};
			\end{pgfonlayer}
			
			\node[squarednode](label)[below of = BackgroundBlue, yshift=-1cm, xshift=-1cm]{flatMapToPair};
			
			\draw (BackgroundPurple.south) -- (label.north);
			\draw (BackgroundBlue.south) -- (label.north);
			\draw (BackgroundOrange.north) -- (label.south);
			\draw (BackgroundGreen.north) -- (label.south);
			
			\draw (label.east) to [out=0, in=200](BackgroundPurpleLarge.west);
			\draw (label.east) -- (BackgroundBlueLarge.west);
			\draw (label.east) to [out=0, in=150] (BackgroundGreenLarge.west);
			
		\end{tikzpicture}	
		\caption{Division and Replication of sub-matrices. Each sub-matrix of dimension $n$ is divided into four sub-matrices of dimension $n/2$ ($A_{11}$ to $A_{22}$) depicted as four color codes using index reordering. The replication is done using \textit{flatMapToPair} transformation. Each sub-matrix has been replicated using algorithm \ref{strassen-algo}.}	
		\label{fig:div-rep}
	\end{figure*}

In the divide step, the matrices are divided into $4$ sub-matrices of equal size and the blocks constitutes each sub-matrix contains same $M-Index$ according to the location of the sub-matrix in its parent. The procedure is shown in Fig. \ref{fig:div-rep}. To create seven sub-matrices $M_{1}$ to $M_{7}$, we create $12$ sub-matrices of size $2^{n-1}$ as shown in Algorithm \ref{strassen-algo}. We create $4$ copies of $A_{11}$ and $A_{22}$ and $2$ copies of $A_{12}$ and $A_{21}$ using \textit{flatMapToPair} transformation. Matrix $B$ is divided similarly. \textit{flatMapToPair} takes a single \textit{key-value} pair and generates a list of \textit{key-value} pairs having the same single key. In this case, it takes a single block of any of the two input matrices and returns a list of blocks according to the group it is about to be consumed i.e. $M_{1}$ to $M_{7}$ and the M-index of the recursive tree. The \textit{mat-name} property of the block preserves predecessor sub-matrix name i.e. $A_{11}$ to $B_{22}$.

\begin{algorithm}[!ht]
\SetAlgoLined
\SetKwFunction{procone}{DivNRep}
\SetKwProg{myproc}{Procedure}{}{}
\myproc{\procone{$RDD<Block> A$, $RDD<Block> B$}}{
    \KwResult{$RDD<Block> C$}
    \tcc{Make union of two input RDDs. Each block of the resulting RDD having a tag with string similar to $A|B$, $M_{1|..|7}$, M-index. In the first recursive call the tag is $A|B,M,0$}
    $RDD<Block> AunionB$ = $A$.union($B$)\;    
    \tcc{Map each block to multiple (key, Block) pairs according to the block index. For example, $A_{11}$ is replicated four times. Each key contains string $M_{1|2...|7}$, M-index. Each block contains a tag with string $A_{11|12|21|22}$ or $B_{11|12|21|22}$.}
    $PairRDD<string,Block> firstMap$ = $AunionB$.flatMapToPair()\;
    \tcc{Group the blocks according to the key. For each key this will group blocks with tags that eventually form $M_{1}$ to $M_{7}$.}
    $PairRDD<string, iterable<Block>> group$ = $firstMap$.groupByKey()\;
    \tcc{Add or subtract blocks with the tag start with similar character $(A|B)$ to get the two blocks of RDDs for the next divide phase.}
    $RDD<Block> C$ = $group$.mapToPair()\;
    \KwRet C\;    
}
\caption{Divide and Replication}
\label{alg:div-rep}
\end{algorithm}

\begin{figure}[!ht]
	\centering
		\begin{tikzpicture}[squarednodePurple/.style={rectangle, draw=purple!80, fill=purple!20, thin, minimum size=1em}, 
		squarednodeGreen/.style={rectangle, draw=green!80, fill=green!20, thin, minimum size=1em},
		squarednodeBlue/.style={rectangle, draw=blue!80, fill=blue!20, thin, minimum size=1em},
		squarednodeOrange/.style={rectangle, draw=orange!80, fill=orange!20, thin, minimum size=1em},
		squarednode/.style={rectangle, draw=black!80, fill=white!20, thin, minimum size=1em},
		invsquarednode/.style={rectangle, draw=white!80, fill=white!20, thin, minimum size=1em},
		backgroundBlue/.style={rectangle,fill=blue!50,inner sep=0.1cm,rounded corners=1mm},
		backgroundRed/.style={rectangle,fill=red!50,inner sep=0.1cm,rounded corners=1mm},
		backgroundYellow/.style={rectangle,fill=yellow!50,inner sep=0.1cm,rounded corners=1mm},
		backgroundGreen/.style={rectangle,fill=green!50,inner sep=0.1cm,rounded corners=1mm},
		node distance=0.4cm,->]
		
			\node[squarednode](00){};
			\node[squarednode](10)[below of = 00]{};
			\node[squarednode](20)[below of = 10, yshift = -0.2cm]{};
			\node[squarednode](30)[below of = 20]{};
			\node[squarednode](40)[below of = 30, yshift = -0.2cm]{};
			\node[squarednode](50)[below of = 40]{};
			\node[squarednode](60)[below of = 50, yshift = -0.2cm]{};
			\node[squarednode](70)[below of = 60]{};
			\node[squarednode](80)[below of = 70, yshift = -0.2cm]{};
			\node[squarednode](90)[below of = 80]{};
			\node[squarednode](100)[below of = 90, yshift = -0.2cm]{};
			\node[squarednode](110)[below of = 100]{};
			\node[squarednode](120)[below of = 110, yshift = -0.2cm]{};
			\node[squarednode](130)[below of = 120]{};
			
			\node[squarednode](01)[right of = 00]{};
			\node[squarednode](11)[right of = 10]{};
			\node[squarednode](21)[right of = 20]{};
			\node[squarednode](31)[right of = 30]{};
			\node[squarednode](41)[right of = 40]{};
			\node[squarednode](51)[right of = 50]{};
			\node[squarednode](61)[right of = 60]{};
			\node[squarednode](71)[right of = 70]{};
			\node[squarednode](81)[right of = 80]{};
			\node[squarednode](91)[right of = 90]{};
			\node[squarednode](101)[right of = 100]{};
			\node[squarednode](111)[right of = 110]{};
			\node[squarednode](121)[right of = 120]{};
			\node[squarednode](131)[right of = 130]{};
			
			\node[squarednode](02)[right of = 01, xshift = 0.7cm]{};
			\node[squarednode](12)[below of = 02]{};
			\node[squarednode](22)[below of = 12, yshift=-0.2cm]{};
			\node[squarednode](32)[below of = 22] {};
			\node[squarednode](42)[below of = 32, yshift=-0.2cm]{};
			\node[squarednode](52)[below of = 42] {};
			\node[squarednode](62)[below of = 52, yshift=-0.2cm]{};
			\node[squarednode](72)[below of = 62] {};
			\node[squarednode](82)[below of = 72, yshift=-0.2cm]{};
			\node[squarednode](92)[below of = 82] {};
			\node[squarednode](102)[below of = 92, yshift=-0.2cm]{};
			\node[squarednode](112)[below of = 102] {};
			\node[squarednode](122)[below of = 112, yshift=-0.2cm]{};
			\node[squarednode](132)[below of = 122] {};
			
			\node[squarednode](03)[right of = 02]{};
			\node[squarednode](13)[below of = 03]{};
			\node[squarednode](23)[below of = 13, yshift=-0.2cm]{};
			\node[squarednode](33)[below of = 23] {};
			\node[squarednode](43)[below of = 33, yshift=-0.2cm]{};
			\node[squarednode](53)[below of = 43] {};
			\node[squarednode](63)[below of = 53, yshift=-0.2cm]{};
			\node[squarednode](73)[below of = 63] {};
			\node[squarednode](83)[below of = 73, yshift=-0.2cm]{};
			\node[squarednode](93)[below of = 83] {};
			\node[squarednode](103)[below of = 93, yshift=-0.2cm]{};
			\node[squarednode](113)[below of = 103] {};
			\node[squarednode](123)[below of = 113, yshift=-0.2cm]{};
			\node[squarednode](133)[below of = 123] {};
			
			\node[squarednode](04)[right of = 03, xshift = 1cm]{};
			\node[squarednode](14)[below of = 04]{};
			\node[squarednode](24)[below of = 14, yshift=-0.2cm]{};
			\node[squarednode](34)[below of = 24] {};
			\node[squarednode](44)[below of = 34, yshift=-0.2cm]{};
			\node[squarednode](54)[below of = 44] {};
			\node[squarednode](64)[below of = 54, yshift=-0.2cm]{};
			\node[squarednode](74)[below of = 64] {};
			\node[squarednode](84)[below of = 74, yshift=-0.2cm]{};
			\node[squarednode](94)[below of = 84] {};
			\node[squarednode](104)[below of = 94, yshift=-0.2cm]{};
			\node[squarednode](114)[below of = 104] {};
			\node[squarednode](124)[below of = 114, yshift=-0.2cm]{};
			\node[squarednode](134)[below of = 124] {};
			
			\node[squarednode](05)[right of = 04]{};
			\node[squarednode](15)[below of = 05]{};
			\node[squarednode](25)[below of = 15, yshift=-0.2cm]{};
			\node[squarednode](35)[below of = 25] {};
			\node[squarednode](45)[below of = 35, yshift=-0.2cm]{};
			\node[squarednode](55)[below of = 45] {};
			\node[squarednode](65)[below of = 55, yshift=-0.2cm]{};
			\node[squarednode](75)[below of = 65] {};
			\node[squarednode](85)[below of = 75, yshift=-0.2cm]{};
			\node[squarednode](95)[below of = 85] {};
			\node[squarednode](105)[below of = 95, yshift=-0.2cm]{};
			\node[squarednode](115)[below of = 105] {};
			\node[squarednode](125)[below of = 115, yshift=-0.2cm]{};
			\node[squarednode](135)[below of = 125] {};
			
			\node[squarednode](06)[right of = 05, xshift = 0.7cm]{};
			\node[squarednode](16)[below of = 06]{};
			\node[invsquarednode](26)[below of = 16, yshift=-0.2cm]{};
			\node[invsquarednode](36)[below of = 26] {};
			\node[invsquarednode](46)[below of = 36, yshift=-0.2cm]{};
			\node[invsquarednode](56)[below of = 46] {};
			\node[invsquarednode](66)[below of = 56, yshift=-0.2cm]{};
			\node[invsquarednode](76)[below of = 66] {};
			\node[invsquarednode](86)[below of = 76, yshift=-0.2cm]{};
			\node[invsquarednode](96)[below of = 86] {};
			\node[squarednode](106)[below of = 96, yshift=-0.2cm]{};
			\node[squarednode](116)[below of = 106] {};
			\node[squarednode](126)[below of = 116, yshift=-0.2cm]{};
			\node[squarednode](136)[below of = 126] {};
			
			\node[squarednode](07)[right of = 06]{};
			\node[squarednode](17)[below of = 07]{};
			\node[invsquarednode](27)[below of = 17, yshift=-0.2cm]{};
			\node[invsquarednode](37)[below of = 27] {};
			\node[invsquarednode](47)[below of = 37, yshift=-0.2cm]{};
			\node[invsquarednode](57)[below of = 47] {};
			\node[invsquarednode](67)[below of = 57, yshift=-0.2cm]{};
			\node[invsquarednode](77)[below of = 67] {};
			\node[invsquarednode](87)[below of = 77, yshift=-0.2cm]{};
			\node[invsquarednode](97)[below of = 87] {};
			\node[squarednode](107)[below of = 97, yshift=-0.2cm]{};
			\node[squarednode](117)[below of = 107] {};
			\node[squarednode](127)[below of = 117, yshift=-0.2cm]{};
			\node[squarednode](137)[below of = 127] {};
			
			\begin{pgfonlayer}{background}
				\node [backgroundRed,fit= (00)(01)(10)(11)] (BackgroundRed00) {};
				\node [backgroundRed,fit= (04)(05)(14)(15)] (BackgroundRed02) {};
				\node [backgroundRed,fit= (24)(25)(34)(35)] (BackgroundRed12) {};
				\node [backgroundRed,fit= (40)(41)(50)(51)] (BackgroundRed20) {};
				\node [backgroundYellow,fit= (62)(63)(72)(73)] (BackgroundYellow31) {};
				\node [backgroundRed,fit= (80)(81)(90)(91)] (BackgroundRed40) {};
				\node [backgroundRed,fit= (104)(105)(114)(115)] (BackgroundRed52) {};
				
				\node [backgroundGreen,fit= (02)(03)(12)(13)] (BackgroundGreen01) {};
				\node [backgroundGreen,fit= (06)(07)(16)(17)] (BackgroundGreen03) {};
				\node [backgroundGreen,fit= (22)(23)(32)(33)] (BackgroundGreen11) {};
				\node [backgroundGreen,fit= (44)(45)(54)(55)] (BackgroundGreen22) {};
				\node [backgroundGreen,fit= (60)(61)(70)(71)] (BackgroundGreen30) {};
				\node [backgroundGreen,fit= (84)(85)(94)(95)] (BackgroundGreen42) {};
				\node [backgroundRed,fit= (102)(103)(112)(113)] (BackgroundRed51) {};
				\node [backgroundGreen,fit= (122)(123)(132)(133)] (BackgroundGreen61) {};
				\node [backgroundGreen,fit= (126)(127)(136)(137)] (BackgroundGreen63) {};
				
				\node [backgroundBlue,fit= (42)(43)(52)(53)] (BackgroundBlue21) {};
				\node [backgroundRed,fit= (64)(65)(74)(75)] (BackgroundRed32) {};
				\node [backgroundBlue,fit= (82)(83)(92)(93)] (BackgroundBlue41) {};
				\node [backgroundBlue,fit= (106)(107)(116)(117)] (BackgroundBlue53) {};
				\node [backgroundBlue,fit= (120)(121)(130)(131)] (BackgroundBlue60) {};
				
				\node [backgroundYellow,fit= (20)(21)(30)(31)] (BackgroundYellow10) {};
				\node [backgroundYellow,fit= (100)(101)(110)(111)] (BackgroundYellow50) {};
				\node [backgroundYellow,fit= (124)(125)(134)(135)] (BackgroundYellow62) {};
				
			\end{pgfonlayer}
			
			\node[invsquarednode](eq1)[right of = BackgroundGreen03, xshift = 0.7cm] {$=$};
			\node[invsquarednode](eq2)[below of = eq1, yshift = -0.6cm] {$=$};
			\node[invsquarednode](eq3)[below of = eq2, yshift = -0.6cm] {$=$};
			\node[invsquarednode](eq4)[below of = eq3, yshift = -0.6cm] {$=$};
			\node[invsquarednode](eq5)[below of = eq4, yshift = -0.6cm] {$=$};
			\node[invsquarednode](eq6)[below of = eq5, yshift = -0.6cm] {$=$};
			\node[invsquarednode](eq7)[below of = eq6, yshift = -0.6cm] {$=$};
			
			\node[invsquarednode](m1)[right of = eq1, xshift = 0.5cm] {$M_{1}$};
			\node[invsquarednode](m2)[right of = eq2, xshift = 0.5cm] {$M_{2}$};
			\node[invsquarednode](m3)[right of = eq3, xshift = 0.5cm] {$M_{3}$};
			\node[invsquarednode](m4)[right of = eq4, xshift = 0.5cm] {$M_{4}$};
			\node[invsquarednode](m5)[right of = eq5, xshift = 0.5cm] {$M_{5}$};
			\node[invsquarednode](m6)[right of = eq6, xshift = 0.5cm] {$M_{6}$};
			\node[invsquarednode](m7)[right of = eq7, xshift = 0.5cm] {$M_{7}$};
			
			\node[invsquarednode](open1)[left of = BackgroundRed00, xshift = -0.3cm] {$($};
			\node[invsquarednode](plus1)[right of = BackgroundRed00, xshift = 0.35cm] {$+$};
			\node[invsquarednode](close1)[right of = BackgroundGreen01, xshift = 0.3cm] {$)$};
			\node[invsquarednode](open2)[right of = close1] {$($};
			\node[invsquarednode](plus2)[right of = BackgroundRed02, xshift = 0.35cm] {$+$};
			\node[invsquarednode](close2)[right of = BackgroundGreen03, xshift = 0.3cm] {$)$};
			
			\node[invsquarednode](open3)[left of = BackgroundYellow10, xshift = -0.3cm] {$($};
			\node[invsquarednode](plus3)[right of = BackgroundYellow10, xshift = 0.35cm] {$+$};
			\node[invsquarednode](close3)[right of = BackgroundGreen11, xshift = 0.3cm] {$)$};
			
			\node[invsquarednode](open4)[left of = BackgroundBlue21, xshift = -0.3cm] {$($};
			\node[invsquarednode](plus4)[right of = BackgroundBlue21, xshift = 0.5cm] {$-$};
			\node[invsquarednode](close4)[right of = BackgroundGreen22, xshift = 0.3cm] {$)$};
			
			\node[invsquarednode](open5)[left of = BackgroundYellow31, xshift = -0.3cm] {$($};
			\node[invsquarednode](plus5)[right of = BackgroundYellow31, xshift = 0.5cm] {$-$};
			\node[invsquarednode](close5)[right of = BackgroundRed32, xshift = 0.3cm] {$)$};
			
			\node[invsquarednode](open6)[left of = BackgroundRed40, xshift = -0.3cm] {$($};
			\node[invsquarednode](plus6)[right of = BackgroundRed40, xshift = 0.35cm] {$+$};
			\node[invsquarednode](close6)[right of = BackgroundBlue41, xshift = 0.3cm] {$)$};
			
			\node[invsquarednode](open7)[left of = BackgroundYellow50, xshift = -0.3cm] {$($};
			\node[invsquarednode](plus7)[right of = BackgroundYellow50, xshift = 0.35cm] {$-$};
			\node[invsquarednode](close7)[right of = BackgroundRed51, xshift = 0.3cm] {$)$};
			\node[invsquarednode](open8)[right of = close7] {$($};
			\node[invsquarednode](plus8)[right of = BackgroundRed52, xshift = 0.35cm] {$+$};
			\node[invsquarednode](close8)[right of = BackgroundBlue53, xshift = 0.3cm] {$)$};
			
			\node[invsquarednode](open9)[left of = BackgroundBlue60, xshift = -0.3cm] {$($};
			\node[invsquarednode](plus9)[right of = BackgroundBlue60, xshift = 0.35cm] {$-$};
			\node[invsquarednode](close9)[right of = BackgroundGreen61, xshift = 0.3cm] {$)$};
			\node[invsquarednode](open10)[right of = close9] {$($};
			\node[invsquarednode](plus10)[right of = BackgroundYellow62, xshift = 0.35cm] {$+$};
			\node[invsquarednode](close10)[right of = BackgroundGreen63, xshift = 0.3cm] {$)$};
			
			\end{tikzpicture}	
		\caption{Addition and Subtraction of sub-matrices. This operation is done using a groupByKey and a mapTopair after the state represented in Figure \ref{fig:div-rep}. groupByKey groups the matrix blocks with tags that will form $M_{1}$ to $M_{7}$. mapToPair adds or subtracts corresponding blocks using tags correspond to matrix $A$ or matrix $B$.}	
		\label{fig:4}
	\end{figure}
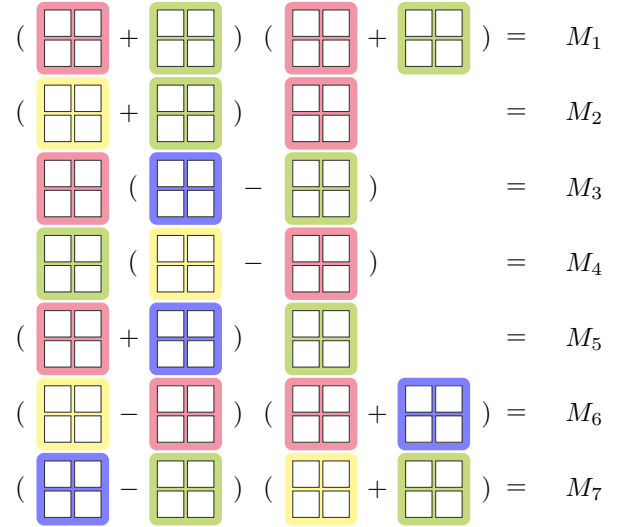

Next, the blocks of similar key ($M_{1}$ to $M_{7}$) are grouped and thus contains the sub-matrices that form $M_{1}$ to $M_{7}$. For adding and subtracting blocks of sub-matrices we use \textit{flapMap} transformation. It takes a \textit{PairRDD} and returns a \textit{RDD}. Here, it takes a list of blocks of $4$ sub-matrices and returns a list of blocks of $2$ sub-matrices. These $2$ sub-matrices are generated using adding or subtracting the corresponding blocks of four sub-matrices as shown in Fig. \ref{fig:4}.

Then we divide the intermediate blocks of the sub-matrices again by recursively calling the Strassen's method. Each time we go down to the leaf of the execution tree, we divide the sub-matrices into smaller sub-matrices of size $2^{n-1}$. This process continues until it reaches the size of a block.

\subsubsection{Multiplication of Block index Matrices}
When the division reaches the size of user-defined block size, the blocks are multiplied using serial matrix multiplication algorithm. This is done using one \textit{mapToPair}, followed by one \textit{groupByKey} and one map function as shown in Algorithm \ref{alg:block-mul}. The \textit{mapToPair} function takes each block and returns a \textit{key-value} pair to group two potential blocks for multiplication. The \textit{groupByKey} groups two blocks and map function returns the multiplication of the two blocks. The keys in the \textit{mapToPair} function are chosen in such a way so that all the leaf index blocks are multiplied in parallel. We transform the leaf node block matrices to Breeze matrices to make the multiplication faster on a single node.

\begin{algorithm}[!ht]
\SetAlgoLined
\SetKwFunction{procone}{MulBlockMat}
\SetKwProg{myproc}{Procedure}{}{}
\myproc{\procone{$RDD<Block> A$, $RDD<Block> B$}}{
    \tcc{Result contains RDD of blocks. Each block is the product of two matrix blocks residing in the same computer.}
    \KwResult{$RDD<Block> C$}
    \tcc{Make union of two input RDDs. Each block of the resulting RDD having a tag with string similar to $A|B, M_{1|2...|7},index$.}
    $RDD<Block> AunionB$ = $A$.union($B$)\;
    \tcc{Map each block to a (key,Block) pair. The key contains string $M_{1|2...|7},index$. Each Block contains a tag with string $A|B$.}
    $PairRDD<string, Block> firstMap$ = $AunionB$.mapToPair()\;
    \tcc{Group the blocks according to the key. For each key, this will group two blocks, one with block tag $A$ and another with $B$.}
    $PairRDD<string, iterable<Block>> group$ = $firstMap$.groupByKey()\;
    \tcc{Multiply two block matrix inside a single computer serially and return each Block to the resulting RDD.}
    $RDD<Block> C$ = $group$.map()\;
    \KwRet C\;    
}
\caption{Block Matrix Multiplication}
\label{alg:block-mul}
\end{algorithm}

\subsubsection{Combining the Sub-matrices}
In this step, the product matrix blocks of seven sub-matrices $M_{1}$ to $M_{7}$ are rearranged to produce a larger matrix. Combination phase occurs when the recursive call to Strassen's procedure returns. In each such return, the size of the matrices becomes $2^{n-1}$ to $2^{n}$. Each such combine step is done in parallel. The \textit{combine} step is shown in Algorithm \ref{alg:combine}.
\vspace{2mm}

This concludes the description of the distributed matrix multiplication algorithm, \textit{Stark}. Next, we theoretically analyze the performance (time complexity) of our algorithm, as well as the baselines.

\begin{algorithm}[!ht]
\SetAlgoLined
\SetKwFunction{procone}{Combine}
\SetKwProg{myproc}{Procedure}{}{}
\myproc{\procone{$RDD<Block> BlockRDD$}}{
    \KwResult{$RDD<Block> C$}
    \tcc{Map each block to (key,Block) pair. Both the key and block mat-name contains string $M_{1|2...|7},index$. indexes are divided by 7 to blocks can be grouped of the same M sub-matrix.}
    $PairRDD<String,Block> firstMap$ = $BlockRDD$.map()\;
    \tcc{Group the blocks that comes from the same M sub-matrix}
    $PairRDD<string, iterable<Block>> group$ = $firstMap$.groupByKey()\;
    \tcc{combine the 7 sub-matrices of size n/2 to a single sub-matrix of size n having the same key}
    $RDD<Block> C$ = $group$.flatMap()\;
    \KwRet C\;    
}
\caption{Combine Phase}
\label{alg:combine}
\end{algorithm}

\section{Performance Analysis}
\label{sec:Performance_Modeling}

In this section, we attempt to estimate the performances of the proposed approach \textit{Stark}, and state-of-the-art approaches \textit{MLLib} and \textit{Marlin}, for distributed matrix multiplication. In this work, we are interested in the \textit{wall clock running time} of the algorithms for varying number of nodes, matrix sizes, and other algorithmic parameters e.g. partition/block sizes. The  wall clock time depends on three independently analyzed quantities: total \textit{computational complexity} of the sub-tasks to be executed (stages in case of spark), total \textit{communication complexity} between executors of different sub-tasks on each of the nodes, and \textit{parallelization factor} of each of the sub-tasks or the total number of processor cores available. Gu et al. \cite{gu2015efficient} also follow a similar paradigm for analysis of \textit{Marlin}.

Later, in section \ref{sec:Experimental-Evaluation}, we compare the theoretically derived estimates of \textit{wall clock time} with empirically observed ones, for validation. The derived performance estimates are also expected to ease the understanding and set tunable parameters for the distributed algorithms. As described before, we consider only square matrices of dimension $2^p$ for all of the derivations. The key input and tunable parameters for the algorithms are:

\begin{itemize}
    \item $n = 2^p$: number of rows or columns in matrix $A$ and $B$ (for square matrix)
    \item $b$ = number of splits or partitions for square matrix
    \item $2^{q} = \frac{n}{b}$ = block size in matrix $A$ and $B$ (used for Strassen's multiplication cost analysis)
    \item $cores$ = Total number of physical cores in the cluster
\end{itemize}
Therefore,
\begin{itemize}
    \item Total number of blocks in matrix $A$ or $B$ = $b^{2}$
    \item $b$ = $2^{p-q}$
\end{itemize}

The following cost analysis has been done conforming to spark execution model which constitutes of two main abstractions --- \textit{Resillient Distributed Dataset} or \textit{RDD} and \textit{Lineage Graph} (which is a DAG (Direct Acyclic Graph) of operations). RDD's are a collection of data items that are split into partitions stored on Hadoop Distributed File System (HDFS).  RDD supports two types of operations --- \textit{transformations} and \textit{actions}. A Spark program implicitly creates a \textit{lineage} which is a logical DAG of transformations that resulted in the RDD. When the driver runs, it converts this logical graph into a physical \textit{execution plan} with a set of \textit{stages} by pipelining the transformations. Then, it creates smaller execution units, referred to as \textit{tasks} under each stage which are bundled up and prepared to be sent to the cluster.

Our analysis follows the execution plan for the programs developed and compared for this paper. This allows us to directly correlate actual wall clock running times with theoretically predicted ones, hence pinpointing the stages which run faster.

\subsection{Performance Analysis of MLLib}
In this section, we derive the cost of the matrix multiplication subroutine of \textit{MLLib} package. As we have used block matrix data structure for \textit{Stark}, the same has been used for experimentation among four choices - \textit{RowMatrix}, \textit{IndexedRowMatrix}, \textit{CoordinateMatrix} and \textit{BlockMatrix}. In the pre-processing step we have transformed the input file to \textit{CoordinateMatrix}, and then converted it into \textit{BlockMatrix}, a data structure synonymous to \textit{Stark's} block matrix structure. While converting to \textit{BlockMatrix}, we provided two parameters - \textit{rowsPerBlock} and \textit{colsPerBlock}. As each block is square, the value of these two parameters are same.

\begin{table*}[!ht]
	\caption{Stagewise performance analysis of MLLib}
	\label{tab:mllibCost}
		\begin{center}
			\begin{tabular}{|c|c|c|c|}
				\hline
				Stage-Step & Computation & Communication & Parallelization Factor \\
				\hline
				Stage 1-flatMap & $b^{3}$ & $NA$ & $min[b^{2}, cores]$ \\
				\hline
				Stage 1-flatMap & $b^{3}$ & $NA$ & $min[b^{2}, cores]$ \\
				\hline
				Stage 3-co-Group & $NA$ & $2\times min[b,cores]\times n^{2}$ & $min[b^{2}, cores]$ \\
				\hline
				Stage 3-flatMap & $b^{3}\times (\frac{n}{b})^{3}$ & $NA$ & $min[b^{2}, cores]$ \\
				\hline
				Stage 4-reduceByKey & $bn^{2}$ & $NA$ & $min[b^{2}, cores]$ \\
				\hline
			\end{tabular}
		\end{center}
\end{table*}

Before multiplying, the scheme partitions the matrix blocks with unique \textit{partitionID} using \textit{GridPartitioner} approach. It partitions the input matrices in a grid structure
and then simulates the multiplication by calculating the destination \textit{partitionIDs} for each block so that only the required blocks is to be shuffled. This cuts the communication cost. For simulation, all the \textit{partitionIDs} needs to be collected at master node. The number of \textit{partitionIDs} for each matrix is $(\frac{n}{b})^{2}$. Therefore, the \textit{communication} cost for this simulation part is

\begin{equation}
Comm_{simulation} = \left( \frac{n}{b} \right)^{2} + \left( \frac{n}{b} \right)^{2} =\frac{2n^{2}}{b^{2}}
\end{equation}

We ignore the computation cost here, because the \textit{groupByKey} and \textit{map} steps are done on a single machine, the cost of which is much less than the overall cost of the approach.

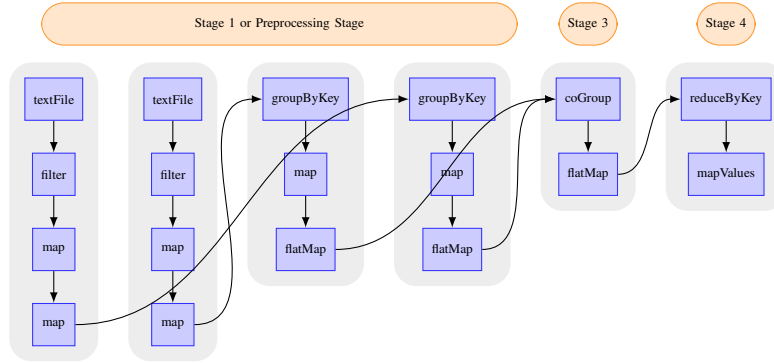
\begin{figure*}[!ht]		
	\centering
	\begin{tikzpicture}[squarednode/.style={rectangle, draw=blue!80, fill=blue!20, thin, minimum size=1.6em}, squarednodeTitle/.style={rectangle, draw=orange!80, fill=orange!20, thin, minimum size=1.6em},
squarednodeTitlePre/.style={rectangle, draw=orange!80, fill=orange!20, thin, minimum size=1.6em, minimum width=18em},	
background/.style={rectangle,fill=gray!14,inner sep=0.2cm,rounded corners=3mm}, node distance=1cm,->]
	
	\node[squarednode](stage11){\tiny textFile};
	\node[squarednode](stage21)[below of = stage11]{\tiny filter};
	\node[squarednode](stage31)[below of = stage21]{\tiny map};
	\node[squarednode](stage41)[below of = stage31]{\tiny map};
	
	\begin{pgfonlayer}{background}
	\node [background,fit= (stage11)(stage21)(stage31)(stage41)] (LargeBox) {};
	\end{pgfonlayer}
	
	\draw (stage11.south) -- (stage21.north);
	\draw (stage21.south) -- (stage31.north);
	\draw (stage31.south) -- (stage41.north);
	
	\node[squarednode](stage12)[right = 0.8cm of stage11]{\tiny textFile};
	\node[squarednode](stage22)[below of = stage12]{\tiny filter};
	\node[squarednode](stage32)[below of = stage22]{\tiny map};
	\node[squarednode](stage42)[below of = stage32]{\tiny map};
	
	\begin{pgfonlayer}{background}
	\node [background,fit= (stage12)(stage22)(stage32)(stage42)] (LargeBox) {};
	\end{pgfonlayer}
	
	\draw (stage12.south) -- (stage22.north);
	\draw (stage22.south) -- (stage32.north);
	\draw (stage32.south) -- (stage42.north);
	
	\node[squarednode](stage13)[right = 0.8cm of stage12]{\tiny groupByKey};
	\node[squarednode](stage23)[below of = stage13]{\tiny map};
	\node[squarednode](stage33)[below of = stage23]{\tiny flatMap};
	
	\begin{pgfonlayer}{background}
	\node [background,fit= (stage13)(stage23)(stage33)] (LargeBox) {};
	\end{pgfonlayer}
	
	\draw (stage13.south) -- (stage23.north);
	\draw (stage23.south) -- (stage33.north);
	
	\node[squarednode](stage14)[right = 0.8cm of stage13]{\tiny groupByKey};
	\node[squarednode](stage24)[below of = stage14]{\tiny map};
	\node[squarednode](stage34)[below of = stage24]{\tiny flatMap};
	
	\begin{pgfonlayer}{background}
	\node [background,fit= (stage14)(stage24)(stage34)] (LargeBox) {};
	\end{pgfonlayer}
	
	\draw (stage14.south) -- (stage24.north);
	\draw (stage24.south) -- (stage34.north);
	
	\node[squarednode](stage15)[right = 0.8cm of stage14]{\tiny coGroup};
	\node[squarednode](stage25)[below of = stage15]{\tiny flatMap};
	
	\begin{pgfonlayer}{background}
	\node [background,fit= (stage15)(stage25)] (LargeBox) {};
	\end{pgfonlayer}
	
	\draw (stage15.south) -- (stage25.north);
	
	\node[squarednode](stage16)[right = 0.8cm of stage15]{\tiny reduceByKey};
	\node[squarednode](stage26)[below of = stage16]{\tiny mapValues};
	
	\begin{pgfonlayer}{background}
	\node [background,fit= (stage16)(stage26)] (LargeBox) {};
	\end{pgfonlayer}
	
	\draw (stage16.south) -- (stage26.north);
	
	\draw (stage41.east) to [out=0, in=180](stage14.west);
	\draw (stage42.east) to [out=0, in=180](stage13.west);
	\draw (stage33.east) to [out=0, in=180](stage15.west);
	\draw (stage34.east) to [out=0, in=180](stage15.west);
	\draw (stage25.east) to [out=0, in=180](stage16.west);
	
	\node[squarednodeTitlePre](stage26)[above of = stage11, rounded corners=3mm, xshift=3cm]{\tiny Stage 1 or Preprocessing Stage};
	\node[squarednodeTitle](stage26)[above of = stage15, rounded corners=3mm]{\tiny Stage 3};
	\node[squarednodeTitle](stage26)[above of = stage16, rounded corners=3mm]{\tiny Stage 4};
	
	\end{tikzpicture}		
	\caption{Execution Plan for \textit{MLLib}}
	\label{fig:mllib-lineage}
\end{figure*}

After simulating, two \textit{flatMap} steps are executed for two input matrices. These are shown in \textit{Stage 1} of the execution plan of \textit{MLLib} in Fig. \ref{fig:mllib-lineage}. The \textit{flatMap} is the actual block replication step where one block of $A$ is taken at a time and replicated as many times as it needs to be multiplied by other blocks of $B$. This value is the number of \textit{partitionIDs} of the blocks of $B$, the input block should multiply with, which is equal to $b$. Therefore, total number of input blocks are $b^{3}$ and computation cost of \textit{Stage 1} is

\begin{equation}
Comp_{Stage1}=2b^{3}
\end{equation}

The parallelization factor for these two steps is

\begin{equation}
PF_{Stage1}=b^{2}
\end{equation}

After that, the actual shuffling takes place using co-group in \textit{Stage 3}. It groups the values associated to similar keys for blocks of both the matrices $A$ and $B$. Therefore, the \textit{communication} cost is equal to 

\begin{equation}
Comm_{co-group}=2\times min[b,cores]\times n^{2}
\end{equation}

The \textit{flatMap} step in \textit{Stage 3} computes the block level multiplication of total $b^{3}$ number of blocks, the cost of which is 

\begin{equation}
\label{eq:leaf-MLLib}
Comp_{flatMap}=b^{3}\times \left(\frac{n}{b}\right)^{3}
\end{equation}

The parallelization factor of \textit{Stage 3} proportional to the total number of partitions of the product matrix, which is

\begin{equation}
PF_{Stage3}=b^{2}
\end{equation}

The \textit{reduceByKey} step of \textit{Stage 4} aggregates the multiplication terms in a group and add them. As all the blocks to be added are already in the same partition there is no need to shuffle and therefore only computation cost for addition is needed. The \textit{Computation} cost for this step is

\begin{equation}
Comp_{reduceByKey}=bn^{2}
\end{equation}

and the parallelization factor is same as above

\begin{equation}
PF_{Stage4}=b^{2}
\end{equation}

Therefore, the total cost of \textit{MLLib} is

\scriptsize
\begin{align}
  \begin{aligned}
    Cost_{MLLib} &= Comm_{Simulation}\\
      &\qquad +(Cost_{Stage1}+Cost_{Stage3}+Cost_{Stage4})\\
      &= \frac{2n^{2}}{b^{2}}+\frac{Comp_{Stage1}}{PF_{Stage1}}\\
      &\qquad + \frac{Comm_{Co-group}+Comp_{flatMap}}{PF_{Stage3}}+\frac{Comp_{Stage4}}{PF_{Stage4}}\\
      &= \frac{2n^{2}}{b^{2}}+\frac{2\times b^{3}}{min[b^{2},cores]}\\
      &\qquad +\frac{2\times min[b,cores]\times n^{2}+b^{3}\times (\frac{n}{b})^{3}}{min[b^{2},cores]}+\frac{bn^{2}}{min[b^{2},cores]}\\
      &= \frac{2n^{2}}{b^{2}}+\frac{2b^{3}+n^{3}+bn^{2}}{min[b^{2},cores]}+\frac{2\times min[b,cores]\times n^{2}}{min[b^{2},cores]}
  \end{aligned}
\end{align}

\normalsize

\subsection{Performance Analysis of Marlin}
\begin{figure*}[!ht]	
	\centering
	\begin{tikzpicture}[squarednode/.style={rectangle, draw=blue!80, fill=blue!20, thin, minimum size=1.6em}, squarednodeTitle/.style={rectangle, draw=orange!80, fill=orange!20, thin, minimum size=1.6em},
squarednodeTitlePre/.style={rectangle, draw=orange!80, fill=orange!20, thin, minimum size=1.6em, minimum width=20em},
background/.style={rectangle,fill=gray!14,inner sep=0.2cm,rounded corners=3mm}, node distance=1cm,->]
	
	\node[squarednode](stage11){\tiny textFile};
	\node[squarednode](stage21)[below of = stage11]{\tiny map};
	\node[squarednode](stage31)[below of = stage21]{\tiny mapPartition};
	
	\begin{pgfonlayer}{background}
	\node [background,fit= (stage11)(stage21)(stage31)] (LargeBox) {};
	\end{pgfonlayer}
	
	\draw (stage11.south) -- (stage21.north);
	\draw (stage21.south) -- (stage31.north);
	
	\node[squarednode](stage12)[right = 0.8cm of stage11]{\tiny groupByKey};
	\node[squarednode](stage22)[below of = stage12]{\tiny mapPartition};
	\node[squarednode](stage32)[below of = stage22]{\tiny flatMap};
	
	\begin{pgfonlayer}{background}
	\node [background,fit= (stage12)(stage22)(stage32)] (LargeBox) {};
	\end{pgfonlayer}
	
	\draw (stage12.south) -- (stage22.north);
	\draw (stage22.south) -- (stage32.north);
	
	\node[squarednode](stage13)[right = 0.8cm of stage12]{\tiny textFile};
	\node[squarednode](stage23)[below of = stage13]{\tiny map};
	\node[squarednode](stage33)[below of = stage23]{\tiny mapPartition};
	
	\begin{pgfonlayer}{background}
	\node [background,fit= (stage13)(stage23)(stage33)] (LargeBox) {};
	\end{pgfonlayer}
	
	\draw (stage13.south) -- (stage23.north);
	\draw (stage23.south) -- (stage33.north);
	
	\node[squarednode](stage14)[right = 0.8cm of stage13]{\tiny groupByKey};
	\node[squarednode](stage24)[below of = stage14]{\tiny map};
	\node[squarednode](stage34)[below of = stage24]{\tiny flatMap};
	
	\begin{pgfonlayer}{background}
	\node [background,fit= (stage14)(stage24)(stage34)] (LargeBox) {};
	\end{pgfonlayer}
	
	\draw (stage14.south) -- (stage24.north);
	\draw (stage24.south) -- (stage34.north);
	
	\node[squarednode](stage15)[right = 0.5cm of stage14]{\tiny partitionBy};
	\node[squarednode](stage15D)[right = 0.1cm of stage15]{\tiny partitionBy};
	\node[squarednode](stage25)[below of = stage15D, xshift=-0.8cm]{\tiny join};
	\node[squarednode](stage35)[below of = stage25]{\tiny mapPartition};
	
	\begin{pgfonlayer}{background}
		\node [background,fit= (stage15)(stage15D)(stage25)(stage35)] (LargeBox) {};
	\end{pgfonlayer}
	
	\draw (stage15.south) -- (stage25.north);
	\draw (stage15D.south) -- (stage25.north);
	\draw (stage25.south) -- (stage35.north);
	
	\node[squarednode](stage16)[right = 0.5cm of stage15D]{\tiny reduceByKey};
	
	\begin{pgfonlayer}{background}
		\node [background,fit= (stage16)] (LargeBox) {};
	\end{pgfonlayer}
	
	\draw (stage31.east) to [out=0, in=180](stage12.west);
	\draw (stage32.east) to [out=0, in=180](stage15.west);
	\draw (stage33.east) to [out=0, in=180](stage14.west);
	\draw (stage34.east) to [out=0, in=180](stage15D.west);
	\draw (stage35.east) to [out=0, in=180](stage16.west);
	
	\node[squarednodeTitlePre](stage26)[above of = stage11, rounded corners=3mm, xshift=2.8cm]{\tiny Stage 1 or Preprocessing Stage};
	\node[squarednodeTitle](stage26)[above of = stage15, rounded corners=3mm, xshift=0.8cm]{\tiny Stage 3};
	\node[squarednodeTitle](stage26)[above of = stage16, rounded corners=3mm]{\tiny Stage 4};
	
	\end{tikzpicture}		
	\caption{Execution Plan for \textit{Marlin}}
	\label{fig:Marlin-Lineage}
\end{figure*}
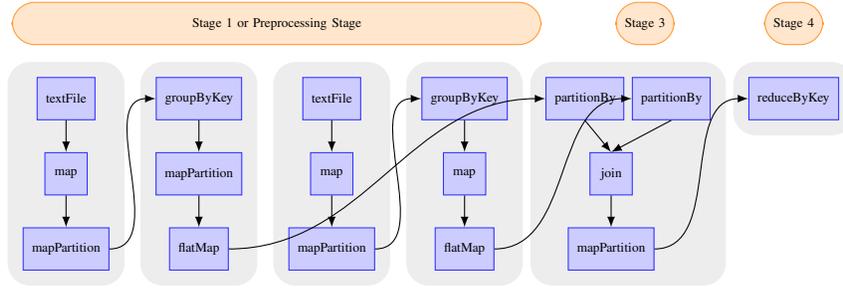

Marlin job execution consists of $6$ stages as shown in Fig. \ref{fig:Marlin-Lineage}. However, the first four stages are part of the preprocessing stage except two \textit{flatMap} transformations. Therefore, two \textit{flatMap} steps and entire \textit{Stage 3} and \textit{Stage 4} are part of the actual execution. The total cost of \textit{Marlin} is given by Lemma \ref{lemma:marlin} and a summary of the cost analysis is tabulated in Table \ref{tab:marlinCost}.

\begin{lemma}
\label{lemma:marlin}
	\textit{Marlin} (referred to as \textit{Block-splitting approach} in \cite{gu2015efficient}) has a complexity in terms of wall clock execution time requirement, where \textbf{$n$} is the matrix dimension, \textbf{$b$} is the number of splits, and \textbf{$cores$} is the actual number of physical cores available in the cluster, as
	
	\scriptsize
	\begin{equation}
	\label{eq:marlin-cost}
		\begin{aligned}
			Cost_{Marlin}&=\frac{4b(b^{2}+n^{2})}{min\left[2b^{2},cores\right]}+\frac{n^{2}(b+n)}{min\left[b^{3},cores\right]}+\frac{bn^{2}}{min\left[b^{2},cores\right]}
		\end{aligned}
	\end{equation}

\normalsize

\end{lemma}

\begin{table*}[!ht]
	\caption{Stagewise cost analysis of Marlin}
	\label{tab:marlinCost}
		\begin{center}
			\begin{tabular}{|c|c|c|c|}
				\hline
				Stage-Step & Computation & Communication & Parallelization Factor \\
				\hline
				Stage 1-flatMap & $2b^{3}$ & $2bn^{2}$ & $min[2b^{2}, cores]$ \\
				\hline
				Stage 1-flatMap & $2b^{3}$ & $2bn^{2}$ & $min[2b^{2}, cores]$ \\
				\hline
				Stage 3-Join & $NA$ & $bn^{2}$ & $min[b^{3}, cores]$ \\
				\hline
				Stage 3-mapPartition & $b^{3}\times (\frac{n}{b})^{3}$ & $bn^{2}$ & $min[b^{3}, cores]$ \\
				\hline
				Stage 4-reduceByKey & $NA$ & $bn^{2}$ & $min[b^{2}, cores]$ \\
				\hline
			\end{tabular}
		\end{center}
\end{table*}

\begin{proof}
To present the proof, we give the $2$ \textit{flatMap} cost of \textit{stage 1}, and total cost of \textit{Stage 3} and \textit{Stage 4}.

\subsubsection{Cost in Stage 1}
Only two transformations in \textit{Stage 1} that contributes to the cost is the \textit{flatMap} steps. Below we derive the cost of \textit{flatMap} step.

\paragraph{Cost in flatMap Step}
In this step, each matrix block is taken as input and a list of matrix blocks is returned. Each block of matrix $A$ returns the number of columns of $B$ blocks and each block of $B$ returns the number of rows of $A$ blocks. Therefore, each block of total $b^{2}$ blocks generates $b$ copies of $A$ blocks and each block of total $b^{2}$ blocks generates $b$ copies of $B$ blocks. As there are $2$ \textit{flatMap} steps, the \textit{computation} cost can be derived as

\begin{equation}
    \begin{aligned}
        Comp_{flatMap}=2 \times (b^{2}\times b)+(b^{2}\times b)=4b^{3}        
    \end{aligned}
\end{equation}

For computing \textit{communication} cost, we need to shuffle $b^{2}$ groups of blocks and each such group consists of $2b$ blocks. Each such block has $\frac{n^{2}}{b^{2}}$ elements. Therefore, total \textit{communication} cost is

\begin{equation}
    \begin{aligned}
    	Comm_{flatMap}=2 \times 2bn^{2}=4bn^{2}
	\end{aligned}
\end{equation}

The parallelization factor depends on how many groups we are processing, which is equal to

\begin{equation}
    \begin{aligned}
        PF_{flatMap}=2b^{2}
    \end{aligned}
\end{equation}

Therefore, total cost in \textit{Stage 1} is

\begin{equation}
    \begin{aligned}
        Cost_{Stage 1}=\frac{4b(b^{2}+n^{2})}{min\left[2b^{2},cores\right]}
    \end{aligned}
\end{equation}

\subsubsection{Cost in Stage 3}
We have two steps in \textit{Stage 3}: \textit{join} and \textit{mapPartition}. We compute the \textit{Communication} cost for \textit{join} and \textit{Computation} and \textit{Communication} cost for \textit{mapPartition}.

\paragraph{Cost in Join Step}
In this step, the output from both the \textit{flatMap} steps are joined, so that related blocks stay together for the next \textit{mapPartition} step of \textit{Stage 3}. It shuffles only one matrix (either $A$ or $B$) through the network. Assuming only shuffling matrix $B$, the cost spending on the network communication can be derived as

\begin{equation}
    \begin{aligned}
        Comm_{Join}
        &=bn^{2}
    \end{aligned}
\end{equation}

The parallelization factor is the number of multiplications done for each block times number of blocks in product matrix. Therefore,

\begin{equation}
    \begin{aligned}
        PF_{Join}
        &=b^{3}
    \end{aligned}
\end{equation}

\paragraph{Cost in mapPartition Step}
In this step the matrix multiplication are conducted locally. To get a single product block, there must be $b^{2}$ block multiplications and there are $b^{2}$ blocks in the product matrix. Each block size is $\frac{n^{2}}{b^{2}}$. Therefore, the computation cost of this step is

\begin{equation}
\label{eq:leaf-Marlin}
    \begin{aligned}
        Comp_{mapPartition}=b^{3}\times (n/b)^{3}
    \end{aligned}
\end{equation}

After the \textit{mapPartition} step the results blocks needs to be written to disks for the next shuffle phase. The \textit{Comm} cost can be derived as

\begin{equation}
    \begin{aligned}
        Comm_{mapPartition}=bn^{2}
    \end{aligned}
\end{equation}

We will not add this value to actual cost as the next shuffle phase incurs the same amount of cost as this. The parallelization factor is the number of multiplications done for each block times number of blocks in product matrix. Therefore,

\begin{equation}
    \begin{aligned}
        PF_{mapPartition}=b^{3}
    \end{aligned}
\end{equation}

Therefore, total cost in \textit{Stage 3} is

\begin{equation}
    \begin{aligned}
        Cost_{Stage 3}=\frac{n^{2}(b+n)}{min\left[b^{3},cores\right]}
    \end{aligned}
\end{equation}

\subsubsection{Cost in Stage 4}
\textit{reduceByKey} is the only step of this stage \textit{reduceByKey} step.

\paragraph{Cost in reduceByKey Step}
In this step, the addition of $b$ blocks is done. Each block has $\frac{n^{2}}{b^{2}}$ elements and \textit{reduceByKey} groups $b$ blocks to get one product block. As there are $b^{2}$ product blocks, the \textit{Communication} cost in this step is

\begin{equation}
    \begin{aligned}
        Comm_{reduceByKey}= b^{2}\times b\times \frac{n^{2}}{b^{2}}=bn^{2}
    \end{aligned}
\end{equation}

In this step, the \textit{parallelization factor} is the number of additions can be done in parallel which is equal to the number of blocks in the product matrix. Therefore,

\begin{equation}
    \begin{aligned}
        PF_{reduceByKey}=b^{2}
    \end{aligned}
\end{equation}

Total cost in \textit{Stage 4} is

\begin{equation}
    \begin{aligned}
        Cost_{Stage 4}=\frac{bn^{2}}{min\left[b^{2},cores\right]}
    \end{aligned}
\end{equation}

Total cost in Marlin algorithm is

\begin{equation}
    \begin{aligned}
        Cost_{Marlin}&=(Cost_{Stage 1}+Cost_{Stage 3}+Cost_{Stage 4}) \\
        &=\frac{4b(b^{2}+n^{2})}{min\left[2b^{2},cores\right]}+\frac{n^{2}(b+n)}{min\left[b^{3},cores\right]}\\&+\frac{bn^{2}}{min\left[b^{2},cores\right]}
    \end{aligned}
\end{equation}

\end{proof}

\subsection{Performance Analysis of Stark}
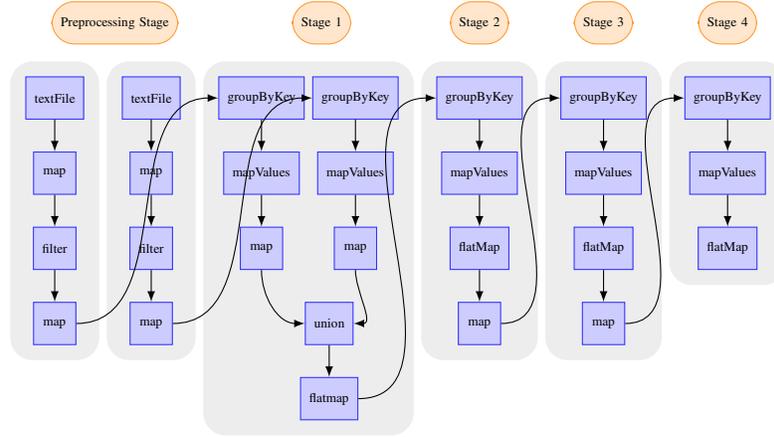
\begin{figure*}[!ht]		
	\centering
	\begin{tikzpicture}[squarednode/.style={rectangle, draw=blue!80, fill=blue!20, thin, minimum size=1.6em}, squarednodeTitle/.style={rectangle, draw=orange!80, fill=orange!20, thin, minimum size=1.6em}, background/.style={rectangle,fill=gray!14,inner sep=0.2cm,rounded corners=3mm}, node distance=1cm,->]
	
	\node[squarednode](stage11){\tiny textFile};
	\node[squarednode](stage21)[below of = stage11]{\tiny map};
	\node[squarednode](stage31)[below of = stage21]{\tiny filter};
	\node[squarednode](stage41)[below of = stage31]{\tiny map};
	
	\begin{pgfonlayer}{background}
	\node [background,fit= (stage11)(stage21)(stage31)(stage41)] (LargeBox) {};
	\end{pgfonlayer}
	
	\draw (stage11.south) -- (stage21.north);
	\draw (stage21.south) -- (stage31.north);
	\draw (stage31.south) -- (stage41.north);
	
	\node[squarednode](stage12)[right = 0.5cm of stage11]{\tiny textFile};
	\node[squarednode](stage22)[below of = stage12]{\tiny map};
	\node[squarednode](stage32)[below of = stage22]{\tiny filter};
	\node[squarednode](stage42)[below of = stage32]{\tiny map};
	
	\begin{pgfonlayer}{background}
	\node [background,fit= (stage12)(stage22)(stage32)(stage42)] (LargeBox) {};
	\end{pgfonlayer}
	
	\draw (stage12.south) -- (stage22.north);
	\draw (stage22.south) -- (stage32.north);
	\draw (stage32.south) -- (stage42.north);
	
	\node[squarednode](stage13)[right = 0.5cm of stage12]{\tiny groupByKey};
	\node[squarednode](stage13D)[right = 0.1cm of stage13]{\tiny groupByKey};
	\node[squarednode](stage23)[below of = stage13]{\tiny mapValues};
	\node[squarednode](stage23D)[below of = stage13D]{\tiny mapValues};
	\node[squarednode](stage33)[below of = stage23]{\tiny map};
	\node[squarednode](stage33D)[below of = stage23D]{\tiny map};
	\node[squarednode](stage43)[below of = stage33, xshift = 0.9cm]{\tiny union};
	\node[squarednode](stage53)[below of = stage43]{\tiny flatmap};
	
	\begin{pgfonlayer}{background}
	\node [background,fit= (stage13)(stage13D)(stage23)(stage23D)(stage33)(stage33D)(stage43)(stage53)] (LargeBox) {};
	\end{pgfonlayer}
	
	\draw (stage13.south) -- (stage23.north);
	\draw (stage13D.south) -- (stage23D.north);
	\draw (stage23.south) -- (stage33.north);
	\draw (stage23D.south) -- (stage33D.north);
	\draw (stage33.south) to [out=270, in=180] (stage43.west);
	\draw (stage33D.south) to [out=270, in=0] (stage43.east);
	\draw (stage43.south) -- (stage53.north);
	
	\node[squarednode](stage14)[right = 0.5cm of stage13D]{\tiny groupByKey};
	\node[squarednode](stage24)[below of = stage14]{\tiny mapValues};
	\node[squarednode](stage34)[below of = stage24]{\tiny flatMap};
	\node[squarednode](stage44)[below of = stage34]{\tiny map};
	
	\begin{pgfonlayer}{background}
	\node [background,fit= (stage14)(stage24)(stage34)(stage44)] (LargeBox) {};
	\end{pgfonlayer}
	
	\draw (stage14.south) -- (stage24.north);
	\draw (stage24.south) -- (stage34.north);
	\draw (stage34.south) -- (stage44.north);
	
	\node[squarednode](stage15)[right = 0.5cm of stage14]{\tiny groupByKey};
	\node[squarednode](stage25)[below of = stage15]{\tiny mapValues};
	\node[squarednode](stage35)[below of = stage25]{\tiny flatMap};
	\node[squarednode](stage45)[below of = stage35]{\tiny map};
	
	\begin{pgfonlayer}{background}
	\node [background,fit= (stage15)(stage25)(stage35)(stage45)] (LargeBox) {};
	\end{pgfonlayer}
	
	\draw (stage15.south) -- (stage25.north);
	\draw (stage25.south) -- (stage35.north);
	\draw (stage35.south) -- (stage45.north);
	
	\node[squarednode](stage16)[right = 0.5cm of stage15]{\tiny groupByKey};
	\node[squarednode](stage26)[below of = stage16]{\tiny mapValues};
	\node[squarednode](stage36)[below of = stage26]{\tiny flatMap};
	
	\begin{pgfonlayer}{background}
	\node [background,fit= (stage16)(stage26)(stage36)] (LargeBox) {};
	\end{pgfonlayer}
	
	\draw (stage16.south) -- (stage26.north);
	\draw (stage26.south) -- (stage36.north);
	
	\node[squarednodeTitle](stage26)[above of = stage11, rounded corners=3mm, align=center, xshift=0.8cm]{\tiny Preprocessing Stage};
	\node[squarednodeTitle](stage26)[above of = stage13, rounded corners=3mm,  xshift=0.8cm]{\tiny Stage 1};
	\node[squarednodeTitle](stage26)[above of = stage14, rounded corners=3mm]{\tiny Stage 2};
	\node[squarednodeTitle](stage26)[above of = stage15, rounded corners=3mm]{\tiny Stage 3};
	\node[squarednodeTitle](stage26)[above of = stage16, rounded corners=3mm]{\tiny Stage 4};
	
	\draw (stage41.east) to [out=0, in=180](stage13.west);
	\draw (stage42.east) to [out=0, in=180](stage13D.west);
	\draw (stage53.east) to [out=0, in=180](stage14.west);
	\draw (stage44.east) to [out=0, in=180](stage15.west);
	\draw (stage45.east) to [out=0, in=180](stage16.west);
	
	\end{tikzpicture}		
	\caption{Execution Plan for \textit{Stark}}
	\label{fig:Stark-Lineage}
\end{figure*}

Unlike \textit{MLLib} and \textit{Marlin}, \textit{Stark} does not posses a constant number of stages as shown in Fig. \ref{fig:Stark-Lineage}. It depends on the number of recursive call to the algorithm. The number of recursive calls are again equal to the logarithm of the number of partitions of the matrix. The total number of stages for a matrix of size of $2^{p}\times 2^{p}$ matrix and $2^{q}\times 2^{q}$ matrix blocks can be obtained using the following equation 

\begin{equation}
    \begin{aligned}
        stages &= 2\log_2(2^{p}/2^{q}) + 2 \\
        &=2\log_22^{p-q} + 2 \\
        &=2(p-q)+2
    \end{aligned}
    \label{eq:Total_Stages}
\end{equation}

For example, as shown in lineage, the number of stages is equal to $4$, when the value of $p-q$ is equal to $1$. To prove it, we divide the algorithm into three main sections - \textit{Divide}, \textit{Multiply} and \textit{Combine}. \textit{Divide} section recursively divides the input matrices into seven sub-matrices and is done by three transformations - \textit{flatMap}, \textit{groupByKey} and \textit{flatMap}. \textit{Multiply} section does the actual multiplication of leaf node matrix blocks. \textit{Combine} section combines the sub-matrices into a single matrix after the recursive call finishes. Both the sections consist of three transformations - \textit{map}, \textit{groupByKey} and \textit{flatMap}. The first \textit{flatMap} transformation of \textit{divide} section and last two transformations of \textit{combine} section require $2$ entire stages resulting $(p-q+1)$ stages each for \textit{divide} and \textit{combine} section. Transfprmations in \textit{multiply} section executes only once. The \textit{map} occupies the last stage of \textit{divide} and \textit{groupByKey} and \textit{flatMap} occupies the first stage of \textit{combine} section. Therefore, total number of stages is equal to $2(p-q)+2$.

\begin{table*}[!ht]
	\caption{Stagewise cost analysis of Stark}
	\label{tab:starkCost}
		\begin{center}
			\begin{tabular}{|l|l|l|l|}
				\hline
				Stage-Step & Computation & Communication & Parallelization Factor \\
				\hline
				\begin{tabular}[c]{@{}l@{}}Stage 1 to Stage p-q\\ - flatMap Divide\end{tabular} & $\frac{8b^{2}}{3}(b^{0.8}-1)$ & $NA$ & $min[(7/4)^{i}(2b^{2}), cores]$ \\
                \hline
				\begin{tabular}[c]{@{}l@{}}Stage 2 to Stage $p-q+1$ \\ - groupByKey Divide\end{tabular} & $NA$ & $\frac{12n^{2}}{5}(b^{1.8}-1)$ & $min[7^{i+1}, cores]$ \\
                \hline
				\begin{tabular}[c]{@{}l@{}}Stage 2 to \\ Stage $p-q+1$ \\ - flatMap Divide\end{tabular} & $\frac{7n^{2}}{3}(b^{0.8}-1)$ & $NA$ & $min[7^{i+1}, cores]$ \\
                \hline
				\begin{tabular}[c]{@{}l@{}}Stage $p-q+1$ \\ - map Leaf\end{tabular} & $2b^{2.8}$ & $NA$ & $min[b^{2.8},cores]$ \\
                \hline
				\begin{tabular}[c]{@{}l@{}}Stage $p-q+2$ \\ - groupByKey Leaf\end{tabular} & $NA$ & $2b^{0.8}n^{2}$ & $min[b^{2.8},cores]$ \\
                \hline
				\begin{tabular}[c]{@{}l@{}}Stage $p-q+2$ \\ - flatMap Leaf\end{tabular} & $b^{2.8}\times (\frac{n}{b})^{3}$ & $NA$ & $min[b^{2.8},cores]$ \\
                \hline
				\begin{tabular}[c]{@{}l@{}}Stage $p-q+2$ to \\ Stage $2p-2q+1$ \\ - map Combine\end{tabular} & $\frac{7b^{2}}{3}(b^{0.8}-1)$ & $NA$ & $min[7^{i+1}, cores]$ \\
                \hline
				\begin{tabular}[c]{@{}l@{}}Stage $p-q+3$ to \\ Stage $2p-2q+2$ \\ - groupByKey Combine\end{tabular} & $NA$ & $\frac{7n^{2}}{3}(b^{0.8}-1)$ & $min[7^{i+1}, cores]$ \\
                \hline
				\begin{tabular}[c]{@{}l@{}}Stage $p-q+3$ to \\ Stage $2p-2q+2$ \\ - flatMap Combine\end{tabular} & $\frac{14n^{2}}{b^{2}}(b^{2.8}-1)$ & $NA$ & $min[7^{i+1}, cores]$ \\
				\hline
			\end{tabular}
		\end{center}
\end{table*}

%
%


\subsubsection{Cost Analysis in Divide Section}
As already stated, \textit{Divide} section consists of three steps - \textit{flatMap}, \textit{groupByKey} and \textit{mapPartition}. Each of the steps occurs $(p-q)$ times.

\paragraph{Cost Analysis of flatMap Step}
There are two types of cost associated with this step - computation cost and \textit{communication} cost. Here, each block is taken as input and according to the tag proper replication is done. As there are $b^{2}$ blocks for each matrix, the computation cost for a single \textit{flatMap} can be derived as

\begin{equation}
    \begin{aligned}
        Comp_{Single flatMap}=2b^{2}
    \end{aligned}
\end{equation}

As the recursion tree going down to the leaf nodes, the number of nodes at each level increases $7$ times, whereas the number of blocks to be processed at each such node decreases by one fourth. Therefore, the total cost for this step for all the stages can be derived as

\begin{equation}
    \begin{aligned}
        Comp_{flatMap}&=\sum_{i=0}^{p-q-1}(7/4)^{i}(2b^{2}) \\        
    \end{aligned}
\end{equation}

The \textit{communication} cost corresponds to the actual number of elements to be shuffled and thus can be derived similarly as

\begin{equation}
    \begin{aligned}
        Comm_{flatMap}&=\sum_{i=0}^{p-q-1}3 \times (7/4)^{i}(2n^{2}) \\        
    \end{aligned}
\end{equation}

As the program progresses the parallelization factor increases as $(7/4)^{i}(2b^{2})$ for $(i = 0, 1, 2, ..., (p-q-1))$.

\paragraph{Cost Analysis of groupByKey Step}
The \textit{communication} cost of this step is

\begin{equation}
    \begin{aligned}
        Comm_{groupByKey}&=\sum_{i=0}^{p-q-1}3 \times (7/2)^{i}(2n^{2}) \\        
    \end{aligned}
\end{equation}

\paragraph{Cost Analysis of flatMap Step}
The computation cost of \textit{flatMap} step can be derived as

\begin{equation}
    \begin{aligned}
        Comp_{flatMap}&=\sum_{i=0}^{p-q-1}(7/2)^{i+1}(2b^{2}) \\        
    \end{aligned}
\end{equation}

As the program progresses the parallelization factor increases as $(7)^{i}$ for $(i = 1, 2, ..., (p-q))$.

\subsubsection{Cost Analysis of Leaf Node Multiplication Section}
There are three steps for this section - \textit{mapPartition}, \textit{groupByKey} and \textit{map}. Each one occurs only once. In the first two steps blocks with similar tags are shuffled and therefore, the \textit{communication} cost of \textit{mapPartition} step can be given as

\small
\begin{equation}
    \begin{aligned}
        Comm_{mapPartition}&=7^{p-q}\times 2\left(\frac{n}{b}\right)^{2}=2\times b^{2.8}\times \left(\frac{n}{b}\right)^{2}      
    \end{aligned}
\end{equation}

\normalsize
and the \textit{communication} cost for \textit{groupByKey} step is

\small
\begin{equation}
    \begin{aligned}
        Comm_{groupByKey}&=7^{p-q}\times 2\left(\frac{n}{b}\right)^{2}=2\times b^{2.8}\times (\frac{n}{b})^{2}          
    \end{aligned}
\end{equation}

\normalsize
The \textit{computation} cost for \textit{map} step is 

\begin{equation}
\label{eq:leaf-Stark}
    \begin{aligned}
        Comp_{map}&=7^{p-q}\left(\frac{n}{b}\right)^{3}=b^{2.8}\times \left(\frac{n}{b}\right)^{3}    
    \end{aligned}
\end{equation}

where actual single node multiplications are done. The parallelization factor is $7^{p-q}$ or $b^{2.8}$.

\subsubsection{Cost Analysis of Combine Section}
In the \textit{mapToPair} step, each block is mapped to a key that denotes the level of the blocks one level up in the recursion tree. There are $7^{i}$ number of different tags at each level $i$ and the number of blocks at level $i$ is $b^{2}/4^{i}$. Therefore, The \textit{computation} cost for \textit{mapToPair} step is

\begin{equation}
    \begin{aligned}
        Comp_{mapPartition}&=\sum_{i=p-q-1}^{0}(7/4)^{i+1}(b^{2}) \\        
    \end{aligned}
\end{equation}

\textit{communication} cost corresponds to saving all these blocks to the disk, which is 

\begin{equation}
\begin{aligned}
Comm_{mapPartition}&=\sum_{i=p-q-1}^{0}(7/4)^{i+1}(n^{2}) \\        
\end{aligned}
\end{equation}

and parallelization factor is $(7)^{i}$ for $(i = (p-q), (p-q-1), ..., 1$.

The \textit{communication} cost for \textit{groupByKey} step is to grouping all the related blocks having same key which is

\begin{equation}
    \begin{aligned}
        Comm_{groupByKey}&=\sum_{i=p-q-1}^{0}(7/4)^{i+1}(n^{2}) \\        
    \end{aligned}
\end{equation}

and the parallelization factor is same as before. 

In the \textit{flatMap} step, blocks with tags $M_{1}$ to $M_{7}$ are transformed into $C_{11}$ to $C_{22}$ sub-matrix blocks. At level $i$, there are $12$ additions of blocks of size $\frac{n}{b}$ where number of groups is $7^{i}$. Therefore, the \textit{computation} cost for \textit{flatMap} step is

\begin{equation}
    \begin{aligned}
       Comp_{flatMap}&=\sum_{i=p-q-1}^{0}7^{i+1}(12\times (\frac{n}{b})^{2}) \\        
    \end{aligned}
\end{equation}

Total cost for \textit{Stage 1}

\begin{equation}
    \begin{aligned}
       Cost_{Stage 1}&=\frac{6n^{2}}{min[1, cores]}        
    \end{aligned}
\end{equation}

Therefore, total Cost for \textit{Stage 2} to \textit{Stage $(p-q)$}

\scriptsize
\begin{equation}
    \begin{aligned}
       Cost_{Stage 2 to (p-q)}&=\sum_{i=0}^{p-q-2}\frac{3.(7/2)^{i}.(2n^{2})}{min[(7/4)^{i}\times 2b^{2}, cores]}\\&+\sum_{i=0}^{p-q-1}\frac{3.(7/2)^{i}.(2n^{2})}{min[7^{i+1}, cores]} \\
       &+\sum_{i=1}^{p-q}\frac{(7/2)^{i}(2b^{2})}{min[7^{i+1}, cores]}+\frac{(7/4)^{p-q}(2n^{2})}{min[7^{p-q}, cores]}
    \end{aligned}
\end{equation}

\normalsize
Total Cost for \textit{Stage (p-q+1)}

\scriptsize
\begin{equation}
    \begin{aligned}
       Cost_{Stage (p-q+1)}&=\frac{2\times b^{2.8}\times (\frac{n}{b})^{2}}{min[b^{2.8}, cores]}+\frac{b^{2.8}(\frac{n}{b})^{3}}{min[b^{2.8}, cores]}\\&+\frac{(7/4)^{p-q}(b^{2})}{min[7^{i}, cores]}
    \end{aligned}
\end{equation}

\normalsize
Total cost for \textit{Stage (p-q+1)} to \textit{Stage (2(p-q)+1)}

\scriptsize
\begin{equation}
    \begin{aligned}
       Cost_{Stage (p-q+2) to (2(p-q)+1)}&=\sum_{i=p-q-1}^{1}\frac{(7/4)^{i}(n^{2})}{min[7^{i}, cores]}\\&+\sum_{i=p-q-1}^{2}\frac{7^{i}(12\times (\frac{n}{b})^{2})}{min[7^{i}, cores]}\\&+\sum_{i=p-q-1}^{1}\frac{(7/4)^{i}(b^{2})}{min[7^{i}, cores]}
    \end{aligned}
\end{equation}

\normalsize
Total cost for \textit{Stage 2(p-q)+2}

\small
\begin{equation}
    \begin{aligned}
       Cost_{Stage (2(p-q)+2)}&=\frac{(7/4)n^{2}}{min[7, cores]}+\frac{7(12\times (\frac{n}{b})^{2})}{min[7, cores]}
    \end{aligned}
\end{equation}

\normalsize

%

\section{Experiments}
\label{sec:Experimental-Evaluation}
In this section, we report results from experiments performed to evaluate the performance of \textit{Stark}, and compare it with that of \textit{Marlin} and \textit{MLLib}. 
First, in section \ref{sec:comparision-with-distributed-systems}, we compare the fastest possible wall clock time, of the three algorithms for different input matrix sizes. Secondly, in section \ref{sec:variation-with-partition-size}, we conduct a series of experiments to individually evaluate the effect of partition size and the matrix size of each competing approaches. At last, we evaluate the scalability of \textit{Stark}.

\subsection{Test Setup}

\begin{table}
	\caption{Test setup components specifications}
	\label{tab:test-setup}
		\begin{center}
			\begin{tabular}{|c|c|c|}
				\hline
				Component Name & Component Size & Specification \\
				\hline
				Processor & 2 & Intel Xeon 2.60 GHz \\
				\hline
				Core & 6 per processor & NA \\
				\hline
				Physical Memory & 132 GB & NA \\
				\hline
				Ethernet & 14 Gb/s & Infini Band \\
				\hline
				OS & NA & CentOS 5 \\
				\hline
				File System & NA\footnotemark[1] & Ext3 \\
				\hline
				Apache Spark & NA & 1.6.0 \\
				\hline
				Apache Hadoop & NA & 2.6.0 \\
				\hline
				Java & NA & 1.7.0 update 79 \\
				\hline
			\end{tabular}
		\end{center}
\end{table}
\footnotetext[1]{Not Applicable}

All the experiments are carried out on a dedicated cluster of $3$ nodes. Software and hardware specifications are summarized in Table \ref{tab:test-setup}. 
For block-level multiplications \textit{Stark} uses Breeze, a single node low-level linear algebra library. Breeze provides Java APIs to \textit{Stark} through Spark, and it calls C/Fortran-implemented native library, such as BLAS, to execute the linear algebra computation through the Java Native Interface (JNI). We have tested the algorithms on matrices with increasing cardinality from $(16 \times 16)$ to $(16384 \times 16384)$. All of these test matrices have been generated randomly using Java Random class. The elements of the matrices are of double-precision $64$-bit IEEE $754$ floating point type.

\subsubsection{Resource Utilization Plan}
While running the jobs in the cluster, we customize three parameters: the number of executors, the executor memory, and the executor cores. We wanted a fair comparison among the competing approaches and therefore, we ensured jobs should not experience \textit{thrashing} and none of the cases tasks should fail and jobs had to be restarted. So, we restricted ourselves to choose the parameter values which provide good utilization of cluster resources and mitigating the chance of task failures. By experimentation, we found that keeping executor memory as $50$ GB ensures successful execution of jobs without \textit{``out of memory''} error or any task failures for all the competing approaches. This includes the small amount of overhead to determine the full request to YARN for each executor which is equal to $3.5$ GB. Therefore, the executor memory is $46.5$ GB. Though the physical memory of each node is $132$ GB, we keep only $100$ GB as YARN resource allocated memory for each node. Therefore, the total physical memory for job execution is $100$ GB resulting $2$ executors per node and a total $6$ executors.  $1$ executor is needed for application manager in YARN and we reserve $1$ core for the operating system and Hadoop daemons. Therefore, the available total number of core is $11$ and total executors are $5$. This leaves $5$ cores for each executor. We used these values of the run-time resource parameters in all the experiments except the scalability test, where we have tested the approach with a varied number of executors. The resource utilization plan is summarized in Table \ref{tab:resourceUtil}.

\begin{table}
	\caption{Resource Utilization Plan for \textit{MLLib}, \textit{Marlin}, and \textit{Stark}}
	\label{tab:resourceUtil}
		\begin{center}
			\begin{tabular}{|c|c|c|}
				\hline
				Component Name & Component Size \\
				\hline
				Executors & 5 \\
				\hline
				Executor Cores & 5 \\
				\hline
				Executor Memory & 50GB \\
				\hline
				YARN Memory & 100GB \\
				\hline
			\end{tabular}
		\end{center}
\end{table}

\subsection{Practical Efficiency of Stark}
\label{sec:comparision-with-distributed-systems}
In this section, we demonstrate the practical utility of \textit{Stark} compare to distributed systems as well as single node optimized matrix multiplication approaches. 

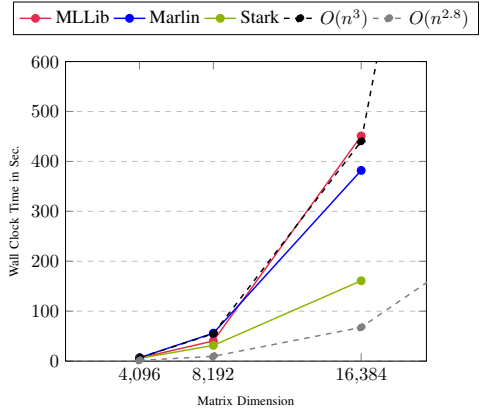
\begin{figure}
	\centering		
		\begin{tikzpicture}[scale=0.7]
		\begin{axis}[
		tick label style={/pgf/number format/fixed},
		scaled ticks=false,
		xlabel={},
		ylabel={},
		xmin=0, xmax=20000,
		ymin=0, ymax=600,
		xtick={4096,8192,16384},
		ytick={0,100,200,300,400,500,600},
		every axis plot/.append style={thick},
		xlabel={\scriptsize Matrix Dimension},
		ylabel={\scriptsize Wall Clock Time in Sec.},
		legend style={at={(0.5,1.2)},
		anchor=north,legend columns=-1},
		ymajorgrids=true,
		grid style=dashed,
		]
		
		\addplot[
		color=red,
		mark=*,
		]
		coordinates {
			(4096,6.1)(8192,40.4)(16384,451)
		};
		
		\addplot[
		color=blue,
		mark=*,
		]
		coordinates {
			(4096,6.8)(8192,56)(16384,382)
		};
	
		\addplot[
		color=green,
		mark=*,
		]
		coordinates {
			(4096,5.5)(8192,31.5)(16384,161)
		};
		
		\addplot[
		color=black,
		dashed,
		mark=*,		
		]
		coordinates {
			(4096,6.8)(8192,55)(16384,440)(32768,3518)
		};
		
		\addplot[
		color=gray,
		dashed,
		mark=*,		
		]
		coordinates {
			(4096,1.38)(8192,9.66)(16384,67.59)(32768,473)
		};
	
		\legend{MLLib, Marlin, Stark, $O(n^{3})$, $O(n^{2.8})$}

		\end{axis}
		\end{tikzpicture}
	\caption{Fastest running time of three systems among different block sizes. Fastest running time occurs 1. at partition size $16$ for all matrix sizes for MLLib, 2. at partition size $4$, $8$, and $16$ for matrix size $4096$, $8192$ and $16384$ respectively for Marlin, and 3. at partition size $4$, $16$, and $16$ for matrix size $4096$, $8192$, and $16384$ respectively for Stark.}\label{fig:Fastest-Running-Time}
\end{figure}

\paragraph{\textbf{Comparison with state-of-the-art distributed systems:}}
In this experiment, we examine the performance of Stark with other Spark based distributed matrix multiplication approaches i.e. \textit{Marlin} and \textit{MLLib}. We report the running time of the competing approaches with increasing matrix size. We take the best wall clock time (fastest) among all the running time taken for different block sizes. The experimental results are shown in Fig. \ref{fig:Fastest-Running-Time}. It can be seen that, \textit{Stark} takes the minimum amount of time for all matrix dimensions, followed by \textit{Marlin}. \textit{MLLib} takes most time. Also, as expected the wall clock execution time increases with the matrix dimension, non-linearly (roughly as $O(n^{2.9})$).
 Also, the gap in wall clock execution time between both \textit{Stark} and \textit{Marlin}, as well as \textit{Marlin} and \textit{MLLib} increases monotonically with input matrix dimension.

\paragraph{\textbf{Comparison with state-of-the-art single node systems}}
The second experiment (Table \ref{tab:CompWithSingleNode}) examines how the runtime improves if we use Stark on the cluster compared to the matrix multiplication on a single node having the similar configuration of a single node in the cluster. We report the fastest wall clock execution times for each of the methods and matrix sizes. The intention of this experiment is to demonstrate the performance of Stark compared to highly optimized linear algebra libraries, e.g. Colt \cite{colt} and JBlas \cite{jblas}. 

We use the parallel version of Colt library, named ParallelColt \cite{wendykier2010parallel} which uses threads automatically when computations are done on a machine with multiple CPUs. 
We also report execution times two other variations of the single node serial matrix multiplication algorithm: the three loop na\"{i}ve approach and the single node Strassen's matrix multiplication algorithm. We use \textit{NA} when the wall clock time is more than reasonable time i.e. more than $1$ hour. 
The results show that initially up to matrix size $(2048\times 2048)$ \textit{Stark} is slower than one other single node algorithm. 

While the comparison here is not fair, since Stark uses much more resources, we show that there is a substantial gain in wall clock time, over the single node parallel options currently available, thus justifying a distributed solution.

\begin{table*}
\caption{Performance comparison among five systems with increasing matrix sizes (The unit of wall clock time is second)}
\label{tab:CompWithSingleNode}
\begin{center}
\begin{tabular}{|c|c|c|c|c|c|}
\hline
    Matrix & Serial Naive & Serial Strassen & Colt & JBlas & Stark (25 executor cores) \\
 	\hline
    $512 \times 512$ & $<1$ & $<1$ & $<1$ & $<1$ & 5 \\
    \hline
    $1024 \times 1024$ & 15 & 2 & 1 & $<1$ & 6 \\
    \hline
    $2048 \times 2048$ & 177 & 14 & 13 & 2 & 9 \\
    \hline
    $4096 \times 4096$ & 2112 & 100 & 135 & 16 & 6.2 \\
    \hline
    $8192 \times 8192$ & NA & 394 & 1163 & 119 & 28.8 \\
    \hline
    $16384 \times 16384$ & NA & 2453 & NA & 862 & 161 \\
    \hline
\end{tabular}
\end{center}
\end{table*}


\begin{figure*}
\centering
	\begin{subfigure}[t]{.3\textwidth}
		\begin{tikzpicture}[scale=1]
		\begin{axis}[
		width=\textwidth,
		xmin=0,
		xmax=33,
		xtick={2,4,8,16,32},
		ytick={50,100,150,200,250},
		x tick label style={/pgf/number format/1000 sep=},
		xlabel={\tiny Partition Size},
		ylabel={\tiny Wall Clock Time in Sec.},
		legend style={at={(0.5,1.2)},
		anchor=north,legend columns=-1},		
		]
		\addplot coordinates {(2,28.2) (4,9.1) (8,7.2) (16,6.1) (32,7.3)(64,8.2)};
		\addplot coordinates {(2,14) (4,6.8) (8,9) (16,28) (32,240)};
		\addplot coordinates {(2,13.8) (4,8) (8,6.2) (16,11.1)(32,22)(64,46)};
		\legend{MLLib, Marlin, Stark}
		\end{axis}
		\end{tikzpicture}
	\end{subfigure}
	\begin{subfigure}[t]{.3\textwidth}
		\begin{tikzpicture}[scale=1]
		\begin{axis}[
		width=\textwidth,
		xmin=0,
		xmax=33,
		xtick={2,4,8,16,32},
		ytick={50,100,150,200,250,300,350},
		x tick label style={/pgf/number format/1000 sep=},
		xlabel={\tiny Partition Size},
		ylabel={\tiny Wall Clock Time in Sec.},
		legend style={at={(0.5,1.2)},
		anchor=north,legend columns=-1},
		]
		\addplot coordinates {(2,217) (4,90.5) (8,66.5) (16,40.4) (32,51.4)};
		\addplot coordinates {(2,101) (4,81) (8,56) (16,63) (32, 246)};
		\addplot coordinates {(2,195) (4,59) (8,33) (16,28.8)(32,71)};
		\legend{MLLib, Marlin, Stark}
		\end{axis}
		\end{tikzpicture}
	\end{subfigure}%
	\begin{subfigure}[t]{.3\textwidth}
		\begin{tikzpicture}[scale=1]
		\begin{axis}[
		width=\textwidth,
		xmin=0,
		xmax=33,
		xtick={4,8,16,32},
		ytick={200,400,600,800},
		x tick label style={/pgf/number format/1000 sep=},
		xlabel={\tiny Partition Size},
		ylabel={\tiny Wall Clock Time in Sec.},
		legend style={at={(0.5,1.2)},
		anchor=north,legend columns=-1},
		]
		\addplot coordinates {(4,722) (8,518) (16,475)(32,451)
		};
		\addplot coordinates {(4,605) (8,521) (16,382)(32,522) };
		\addplot coordinates {(4,541) (8,278) (16,262)(32,376)(64,571)};
		\legend{MLLib, Marlin, Stark}
		\end{axis}
		\end{tikzpicture}
	\end{subfigure}%
	\caption{Comparing running time of MLLib, Marlin, and Stark for matrix size $(4096\times 4096)$, $(8192\times 8192)$, $(16384\times 16384)$ for increasing partition size. The minimum wall clock execution times for the methods correspond to the fastest running time as shown in Figure \ref{fig:Fastest-Running-Time}.}\label{fig:Running-Time-Size}
\end{figure*}
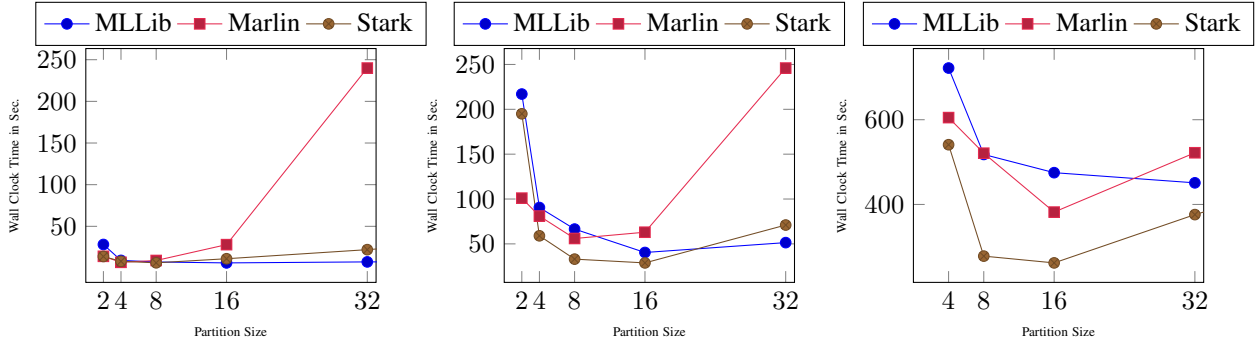

\subsection{Variation with partition size}
\label{sec:variation-with-partition-size}
In this experiment, we examine the performance of \textit{Stark} with \textit{Marlin} and \textit{MLLib} with increasing partition size for each matrix size. We report the wall clock execution time of the approaches when partition size is increased within a matrix size. For each matrix size (from $(4096\times 4096)$ to $(16384\times 16384)$ we increase the partition size until we get an intuitive change in the results as shown in Figure \ref{fig:Running-Time-Size}.

We see that \textit{Stark} takes the minimum amount of wall clock time among all the approaches for almost all the partition sizes and for all the matrix sizes. The costliest part of all the competing approaches is \textit{Stage3}, which contains shuffle and leaf node block matrix multiplication steps. It can be easily verified that the computation cost of \textit{Stark} (as shown in equation \ref{eq:leaf-Stark}) is less than \textit{MLLib} (equation \ref{eq:leaf-MLLib}) and \textit{Marlin} (equation \ref{eq:leaf-Marlin}). This is studied in further detail in section \ref{sec:thVsEx}.

We see that \textit{Stark} takes more time than \textit{MLLib}  at $b=16$ and $b=32$ and matrix size $4096\times 4096$ and $8192\times 8192$. The divide section cost at \textit{Stage 2} of \textit{Stark} depends on $b$ or consequently on the value of $p-q$. Therefore, as the value of $b$ increases, more the number of times divide section executes, resulting in additive cost. This suggests that too many partitions for a small matrix hurt the wall clock execution time.

It can be seen that all the approaches follow a \textit{U} shaped curve i.e. for smaller values of partition size $b$ the wall clock running time is large and as we increase $b$, it takes an optimal wall clock time and again increases for a further increase in $b$. The reason is that, for all the approaches the dominating cost is leaf node multiplications (as given in equation \ref{eq:leaf-MLLib}, \ref{eq:leaf-Marlin} and \ref{eq:leaf-Stark}) and it depends on the parallelization factor or \textit{PF}. The \textit{PF} of all the approaches increases as $b$ increases resulting gradual decreasing of the cost and then it becomes constant as the value of \textit{cores}. After that, the total cost started increasing as other cost increases with increasing $b$.

The gap between \textit{MLLib} and \textit{Stark} decreases as $b$ increases. On the other hand, after optimal point \textit{Stark} line overshoots but \textit{MLLib} got more gradual tendency. This is because, there is a trade-off between the computation cost and communication cost of \textit{Stark}, as we increase the partition size. For a smaller number of partitions, the computation cost is comparatively higher, while communication cost is low due to short height of the recursion tree. On the other hand, large partition size increases communication cost for each level of the recursion tree, without gaining much from parallelization. This suggests that unrolling the recursion to an appropriate depth will result in an optimal gap in performance.

\subsection{Comparison between theoretical and experimental results}
\label{sec:thVsEx}
\begin{figure*}
\centering
	\begin{subfigure}[t]{.3\textwidth}
		\begin{tikzpicture}
		\begin{axis}[
		xmin=0,
		xmax=33,
		xtick={2,4,8,16,32},
		ytick={10,20,30,40,50,60,70},
		x tick label style={/pgf/number format/1000 sep=},
		xlabel={\tiny Partition Size},
		ylabel={\tiny Wall Clock Time in Sec.},
		legend style={at={(0.5,1.5)},anchor=north},
		width=\textwidth,
		yticklabel style = {font=\tiny},
        xticklabel style = {font=\tiny},
		]
		\addplot coordinates {(2,20.5520896) (4,5.97898036) (8,4.902155756) (16,7.054615183)(32,9.481175695)};
		\addlegendentry{Theoretical}
		\addplot coordinates {(2,28.2) (4,9.1) (8,7.2) (16,6.1)(32,7.3)};
		\addlegendentry{Experimantal}
		\end{axis}
		\end{tikzpicture}
		\caption{$4096\times 4096$}
	\end{subfigure}%
	\begin{subfigure}[t]{.3\textwidth}
		\begin{tikzpicture}
		\begin{axis}[
		xmin=0,
		xmax=33,
		xtick={2,4,8,16,32},
		ytick={50,100,150,200,250,300},
		x tick label style={/pgf/number format/1000 sep=},
		xlabel={\tiny Partition Size},
		ylabel={\tiny Wall Clock Time in Sec.},
		legend style={at={(0.5,1.5)},anchor=north},
		width=\textwidth,
		yticklabel style = {font=\tiny},
        xticklabel style = {font=\tiny},
		]
		\addplot coordinates {(2,150.9278351) (4,41.0957906) (8,30.60373918) (16,39.21357603)(32,48.91981119)};
		\addlegendentry{Theoretical}
		\addplot coordinates {(2,217) (4,90.5) (8,66.5) (16,40.4)(32,51.4)(64,78.5)};
		\addlegendentry{Experimantal}
		\end{axis}
		\end{tikzpicture}
		\caption{$8192\times 8192$}
	\end{subfigure}%
	\begin{subfigure}[t]{.3\textwidth}
		\begin{tikzpicture}
		\begin{axis}[
		xmin=0,
		xmax=33,
		xtick={4,8,16,32},
		ytick={150,300,450,600,750},
		x tick label style={/pgf/number format/1000 sep=},
		xlabel={\tiny Partition Size},
		ylabel={\tiny Wall Clock Time in Sec.},
		legend style={at={(0.5,1.5)},anchor=north},
		width=\textwidth,
		yticklabel style = {font=\tiny},
        xticklabel style = {font=\tiny},
		]
		\addplot coordinates {(4,301.8221158) (8,210.3758868) (16,244.8152334) (32,283.6401671)};
		\addlegendentry{Theoretical}
		\addplot coordinates {(4,722) (8,518) (16,475)(32,451)};
		\addlegendentry{Experimantal}
		\end{axis}
		\end{tikzpicture}
		\caption{$16384\times 16384$}
	\end{subfigure}

	\begin{subfigure}[t]{.3\textwidth}
		\begin{tikzpicture}
		\begin{axis}[
		xmin=0,
		xmax=33,
		xtick={2,4,8,16,32},
		ytick={50,100,150,200},
		x tick label style={/pgf/number format/1000 sep=},
		xlabel={\tiny Partition Size},
		ylabel={\tiny Wall Clock Time in Sec.},
		legend style={at={(0.5,1.5)},anchor=north},
		width=\textwidth,
		yticklabel style = {font=\tiny},
        xticklabel style = {font=\tiny},
		]
		\addplot coordinates {(2,12.80101581) (4,5.441522248) (8,8.12822569) (16,13.5076728)(32,24.26657047)};
		\addlegendentry{Theoretical}
		\addplot coordinates {(2,14) (4,6.8) (8,9) (16,28) (32,240)};
		\addlegendentry{Experimantal}
		\end{axis}
		\end{tikzpicture}
		\caption{$4096\times 4096$},
	\end{subfigure}%
	\begin{subfigure}[t]{.3\textwidth}
		\begin{tikzpicture}
		\begin{axis}[
		xmin=0,
		xmax=33,
		xtick={2,4,8,16,32},
		ytick={50,100,150,200,250,300,350},
		x tick label style={/pgf/number format/1000 sep=},
		xlabel={\tiny Partition Size},
		ylabel={\tiny Wall Clock Time in Sec.},
		legend style={at={(0.5,1.5)},anchor=north},
		width=\textwidth,
		yticklabel style = {font=\tiny},
        xticklabel style = {font=\tiny},
		]
		\addplot coordinates {(2,85.5638016) (4,32.76120524) (8,43.50801879) (16,65.02580552)(32,108.0613824)};
		\addlegendentry{Theoretical}
		\addplot coordinates {(2,101) (4,81) (8,56) (16,63) (32,246)};
		\addlegendentry{Experimantal}
		\end{axis}
		\end{tikzpicture}
		\caption{$8192\times 8192$}
	\end{subfigure}%
	\begin{subfigure}[t]{.3\textwidth}
		\begin{tikzpicture}
		\begin{axis}[
		xmin=0,
		xmax=33,
		xtick={2,4,8,16,32},
		ytick={200,400,600,800,1000},
		x tick label style={/pgf/number format/1000 sep=},
		xlabel={\tiny Partition Size},
		ylabel={\tiny Wall Clock Time in Sec.},
		legend style={at={(0.5,1.5)},anchor=north},
		width=\textwidth,
		yticklabel style = {font=\tiny},
        xticklabel style = {font=\tiny},
		]
		\addplot coordinates {(4,617.1331133) (8,219.0057511) (16,261.9930051) (32,520.2064441)};
		\addlegendentry{Theoretical}
		\addplot coordinates {(2,810) (4,605) (8,521) (16,382) (32, 522)};
		\addlegendentry{Experimantal}
		\end{axis}
		\end{tikzpicture}
		\caption{$16384\times 16384$}
	\end{subfigure}

	\begin{subfigure}[t]{.3\textwidth}
		\begin{tikzpicture}
		\begin{axis}[
		xmin=0,
		xmax=33,
		xtick={2,4,8,16,32},
		ytick={20,40,60,80,100},
		x tick label style={/pgf/number format/1000 sep=},
		xlabel={\tiny Partition Size},
		ylabel={\tiny Wall Clock Time in Sec.},
		legend style={at={(0.5,1.5)},anchor=north},
		width=\textwidth,
		yticklabel style = {font=\tiny},
        xticklabel style = {font=\tiny},
		]
		\addplot coordinates {(2,14.0203599) (4,8.496797108) (8,12.06163618) (16,18.53726359)(32,86.25753128)};
		\addlegendentry{Theoretical}
		\addplot coordinates {(2,15) (4,5.5) (8,6.4) (16,12.4)(32,22)(64,46)};
		\addlegendentry{Experimantal}
		\end{axis}
		\end{tikzpicture}
		\caption{$4096\times 4096$}
	\end{subfigure}%
	\begin{subfigure}[t]{.3\textwidth}
		\begin{tikzpicture}
		\begin{axis}[
		xmin=0,
		xmax=33,
		xtick={2,4,8,16,32},
		ytick={30,60,90,120},
		x tick label style={/pgf/number format/1000 sep=},
		xlabel={\tiny Partition Size},
		ylabel={\tiny Wall Clock Time in Sec.},
		legend style={at={(0.5,1.5)},anchor=north},
		width=\textwidth,
		yticklabel style = {font=\tiny},
        xticklabel style = {font=\tiny},
		]
		\addplot coordinates {(2,90.44117797) (4,42.31992834) (8,55.5006159) (16,80.46408837)(32,125.8018366)};
		\addlegendentry{Theoretical}
		\addplot coordinates {(2,97) (4,46) (8,32) (16,31.5)(32,71)(64,117)};
		\addlegendentry{Experimantal}
		\end{axis}
		\end{tikzpicture}
		\caption{$8192\times 8192$}
	\end{subfigure}%
	\begin{subfigure}[t]{.3\textwidth}
		\begin{tikzpicture}
		\begin{axis}[
		xmin=0,
		xmax=33,
		xtick={2,4,8,16,32},
		ytick={150,300,450,600},
		x tick label style={/pgf/number format/1000 sep=},
		xlabel={\tiny Partition Size},
		ylabel={\tiny Wall Clock Time in Sec.},
		legend style={at={(0.5,1.5)},anchor=north},
		width=\textwidth,
		yticklabel style = {font=\tiny},
        xticklabel style = {font=\tiny},
		]
		\addplot coordinates {(2,636.6426188108572)(4,235.9416329) (8,280.035035) (16,372.3766394) (32,547.1877975)};
		\addlegendentry{Theoretical}
		\addplot coordinates {(4,441) (8,253) (16,161) (32,376)};
		\addlegendentry{Experimantal}
		\end{axis}
		\end{tikzpicture}
		\caption{$16384\times 16384$}
	\end{subfigure}
	\caption{Comparing theoretical and experimental running time of \textit{MLLib} (a),(b) and (c), \textit{Marlin} (d),(e) and (f) and \textit{Stark} (g), (h) and (i) for matrix size $(4096\times 4096)$, $(8192\times 8192)$, $(16384\times 16384)$ for increasing partition size}\label{fig:Running-Time-Partition-MLLib}
\end{figure*}
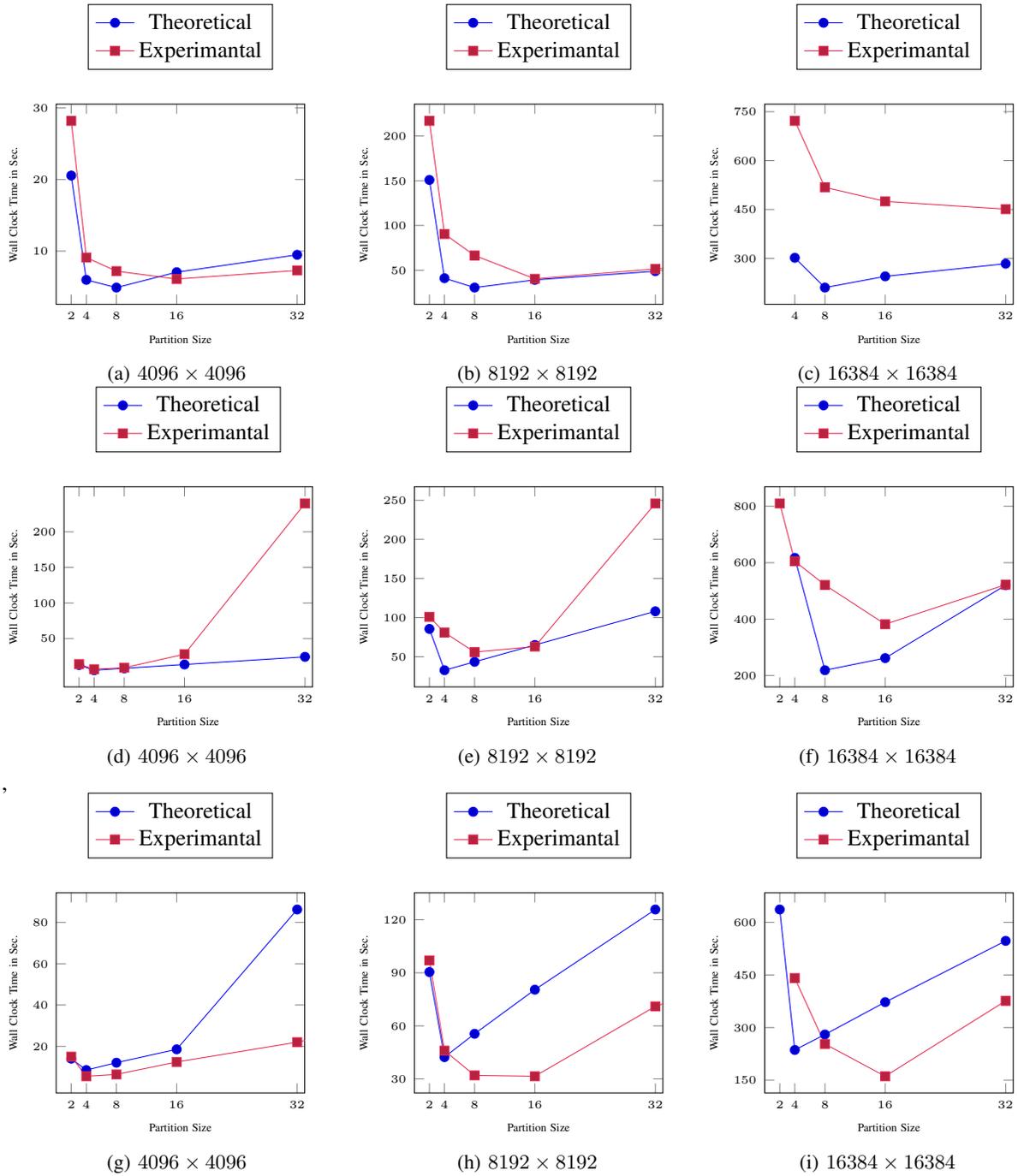

In this experiment, we compare the theoretical cost of all the approaches with the experimental wall clock execution time to validate our theoretical cost analysis. Fig. \ref{fig:Running-Time-Partition-MLLib} shows the comparison for three matrix sizes (from $(4096\times 4096)$ to $(16384\times 16384)$) and for each matrix size with increasing partition size.

As expected, both the theoretical and experimental wall clock execution time for all the three approaches shows a \textit{U} shaped curve with increasing partition size. The reason is that the size of matrix blocks becomes very large when it is partitioned into smaller partitions for very large matrices. As a result, the single node matrix multiplication execution time is found to be large at the beginning which results in overall large execution time. As the partition size increases, the parallelization factor, equal to $min\left[b^{3},cores\right]$ or $min\left[b^{2.8},cores\right]$, increases until it reaches the actual number of physical cores available in the cluster. As a result, the overall cost decreases gradually. After the optimal point is reached, the computation cost stabilizes and communication cost starts to increase as the factor $\left[bn^{2}\right]$ for \textit{MLLib} and \textit{Marlin} and $\left[\sum_{i=0}^{p-q-1}(7/4)^{i}(2b^{2})\right]$ for \textit{Stark}, increases.

\begin{table*}[!ht]
\caption{Theoretical (marked in green) and actual (marked in red) Computation Cost (in milliseconds) For Leaf Node Block Matrix Multiplications for \textit{Marlin} and \textit{Stark} for Partition Size from $2$ to $32$ and Matrix dimension from $4096$ to $16384$}
\label{tab:CompWithSingleNode}
\begin{center}
\begin{tabularx}{\textwidth}{|X|X|X|X|X|X|}
\hline
    \multicolumn{6}{|c|}{Matrix Size: $4096\times 4096$} \\
    \hline
    \multirow{2}{*}{Method} & \multicolumn{5}{|c|}{Number of Partitions} \\ \cline{2-6}
     & 2 & 4 & 8 & 16 & 32 \\
    \hline
    Marlin & 10225 & \cellcolor{green}6578 & \cellcolor{red}5059 & 7012 & 15598 \\   
	\hline
    Stark & 12533 & \cellcolor{green}4715 & 3724 & 3368 & \cellcolor{red}3123\\
    \hline
    \multicolumn{6}{|c|}{Matrix Size: $8192\times 8192$} \\
    \hline
    \multirow{2}{*}{Method} & \multicolumn{5}{|c|}{Number of Partitions} \\ \cline{2-6}
     & 2 & 4 & 8 & 16 & 32 \\
    \hline
    Marlin & 93293 & \cellcolor{green}63050 & 42201 & \cellcolor{red}34777 & 42191 \\    
	\hline
    Stark & 99185 & \cellcolor{green}64715 & 31348 & \cellcolor{red}23945 & 35662 \\
    \hline
    \multicolumn{6}{|c|}{Matrix Size: $16384\times 16384$} \\
    \hline
    \multirow{2}{*}{Method} & \multicolumn{5}{|c|}{Number of Partitions} \\ \cline{2-6}
     & 2 & 4 & 8 & 16 & 32 \\
    \hline
    Marlin & 681401 & \cellcolor{green}433659 & \cellcolor{red}325698 & 335291 & 413648 \\ 
	\hline
    Stark & 680896 &  \cellcolor{green}412706 & 246895 & \cellcolor{red}175346 & 186747 \\
    \hline
\end{tabularx}
\end{center}
\end{table*}

From Fig. \ref{fig:Running-Time-Partition-MLLib} we observe that the minimum actual time and minimum theoretical time occur at different partition sizes, for example in \textit{MLLib} for matrix size $8192\times 8192$ the minimum theoretical time occurs at partition size $8$, whereas minimum practical time occurs at partition size $16$. We found this discrepancy to explainable using computation wall clock time reported in Table \ref{tab:CompWithSingleNode}. These costs have been extracted by caching matrix blocks to be multiplied at the leaf nodes into the executor memory and calculating the execution time for just the transformations used for leaf node block multiplications. For brevity, we report times only for \textit{Marlin} and \textit{Stark} as \textit{MLLib} follows similar pattern as \textit{Marlin}. Green and red cells mark the minimum of theoretical and experimental computation times divided by parallelization factor, respectively.

As observed in Fig. \ref{fig:Running-Time-Partition-MLLib}, the experimental running time minimum comes later than theoretical one. We observe that the shifts in computation cost minimum (Table \ref{tab:CompWithSingleNode}) roughly correspond to the shifts in overall execution time (Fig. \ref{fig:Running-Time-Partition-MLLib}). For example, for \textit{Stark} the minimum computation cost comes at partition size $16$ and so the overall cost. Since we use the breeze library for multiplication of matrix blocks, we attribute this discrepancy to internal optimizations of the breeze package. Hence, we conclude that modulo this discrepancy, the theoretical and experimental wall clock times match, thus justifying our analysis.

\subsection{Stage-wise Comparison}
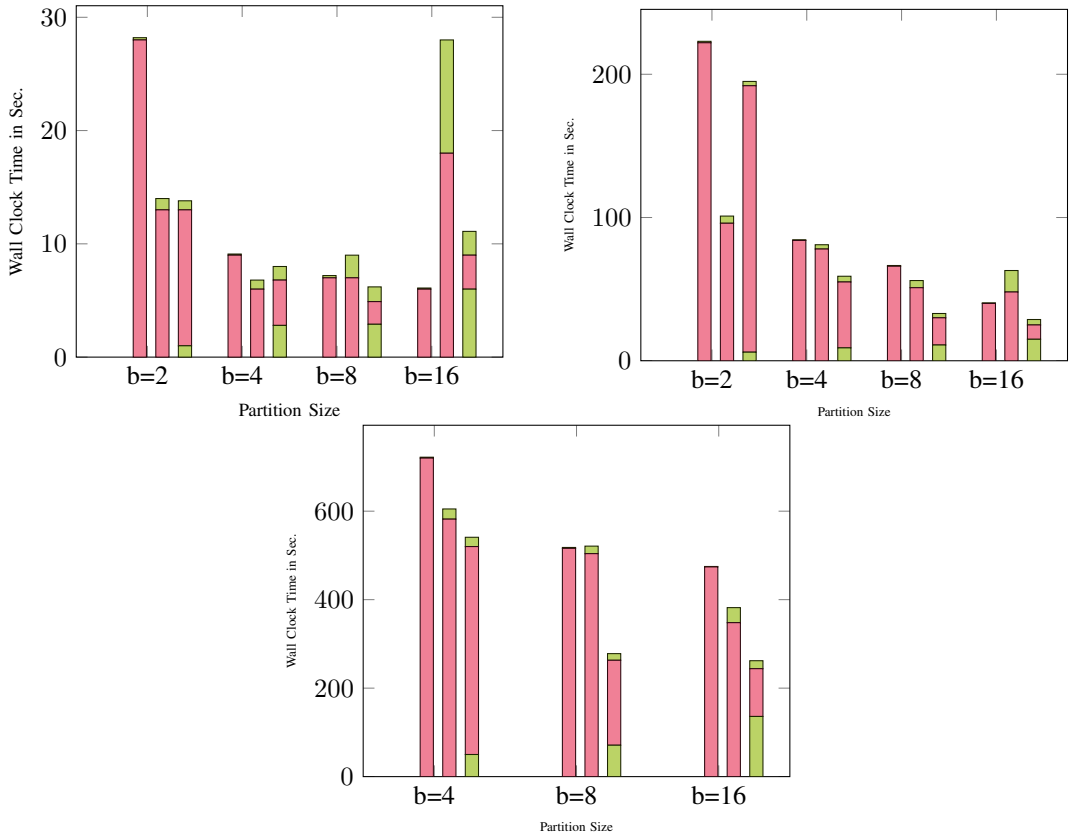
\begin{figure*}[!ht]
\centering
	\begin{subfigure}[t]{.4\textwidth}
\begin{tikzpicture}
\begin{axis}[
width=\textwidth,
ybar stacked,bar width=5,
ymin=0,
xmin=2, xmax=8,
enlarge x limits=0.25,
legend style={at={(1,1)},anchor=north west},
xlabel={\scriptsize Partition Size},
ylabel={\scriptsize Wall Clock Time in Sec.},
xtick={2,4,6,8},
xticklabels={b=2,b=4,b=8,b=16}
]
\addplot +[bar shift=-.1cm][red!20!black,fill=red!60!white] coordinates {(2,28) (4,9) (6,7) (8,6)};
\addplot +[bar shift=-.1cm][green!20!black,fill=green!60!white] coordinates {(2,0.2) (4,0.1) (6,0.2) (8,0.1)};

\resetstackedplots

\addplot +[bar shift=.2cm][red!20!black,fill=red!60!white] coordinates {(2,13) (4,6) (6,7) (8,18)};
\addplot +[bar shift=.2cm][green!20!black,fill=green!60!white] coordinates {(2,1) (4,0.8) (6,2) (8,10)};

\resetstackedplots

\addplot +[bar shift=.5cm][green!20!black,fill=green!60!white] coordinates {(2,1) (4,2.8) (6,2.9) (8,6)};
\addplot +[bar shift=.5cm][red!20!black,fill=red!60!white] coordinates {(2,12) (4,4) (6,2) (8,3)};
\addplot +[bar shift=.5cm][green!20!black,fill=green!60!white] coordinates {(2,0.8) (4,1.2) (6,1.3) (8,2.1)};
\end{axis}
\end{tikzpicture}
\end{subfigure}
\begin{subfigure}[t]{.4\textwidth}
	\begin{tikzpicture}
	\begin{axis}[
	width=\textwidth,
	ybar stacked,bar width=5,
	ymin=0,
	xmin=2, xmax=8,
	enlarge x limits=0.25,
	legend style={at={(1,1)},anchor=north west},
	xlabel={\tiny Partition Size},
	ylabel={\tiny Wall Clock Time in Sec.},
	xtick={2,4,6,8},
	xticklabels={b=2,b=4,b=8,b=16},
	]
	\addplot +[bar shift=-.1cm][red!20!black,fill=red!60!white] coordinates {(2,222) (4,84) (6,66) (8,40)};
	\addplot +[bar shift=-.1cm][green!20!black,fill=green!60!white] coordinates {(2,1) (4,0.4) (6,0.5) (8,0.4)};
	
	\resetstackedplots
	
	\addplot +[bar shift=.2cm][red!20!black,fill=red!60!white] coordinates {(2,96) (4,78) (6,51) (8,48)};
	\addplot +[bar shift=.2cm][green!20!black,fill=green!60!white] coordinates {(2,5) (4,3) (6,5) (8,15)};
	
	\resetstackedplots
	
	\addplot +[bar shift=.5cm][green!20!black,fill=green!60!white] coordinates {(2,6) (4,9) (6,11) (8,15)};
	\addplot +[bar shift=.5cm][red!20!black,fill=red!60!white] coordinates {(2,186) (4,46) (6,19) (8,10)};
	\addplot +[bar shift=.5cm][green!20!black,fill=green!60!white] coordinates {(2,3) (4,4) (6,3) (8,3.8)};
	\end{axis}
	\end{tikzpicture}
\end{subfigure}%
\\
\begin{subfigure}[t]{.4\textwidth}
	\begin{tikzpicture}
	\begin{axis}[
	width=\textwidth,
	ybar stacked,bar width=5,
	xtick={4,8,16},
	ymin=0,
	xmin=4, xmax=8,
	enlarge x limits=0.25,
	legend style={at={(1,1)},anchor=north west},
	xlabel={\tiny Partition Size},
	ylabel={\tiny Wall Clock Time in Sec.},
	xtick={4,6,8},
	xticklabels={b=4,b=8,b=16},
	]
	\addplot +[bar shift=-.1cm][red!20!black,fill=red!60!white] coordinates {(4,720) (6,516) (8,474)};
	\addplot +[bar shift=-.1cm][green!20!black,fill=green!60!white] coordinates {(4,2) (6,2) (8,1)};
	
	\resetthreestackedplots
	
	\addplot +[bar shift=.2cm][red!20!black,fill=red!60!white] coordinates {(4,582) (6,504) (8,348)};
	\addplot +[bar shift=.2cm][green!20!black,fill=green!60!white] coordinates {(4,23) (6,17) (8,34)};
	
	\resetthreestackedplots
	
	\addplot +[bar shift=.5cm][green!20!black,fill=green!60!white] coordinates {(4,50) (6,71) (8,136)};
	\addplot +[bar shift=.5cm][red!20!black,fill=red!60!white] coordinates {(4,470) (6,192) (8,108)};
	\addplot +[bar shift=.5cm][green!20!black,fill=green!60!white] coordinates {(4,21) (6,15) (8,18)};

	\end{axis}
	\end{tikzpicture}
\end{subfigure}%
\caption{Comparing step-wise running time among MLLib, Marlin and Stark for matrix size $(4096\times 4096)$, $(8192\times 8192)$, $(16384\times 16384)$ and for increasing partition size. Here red bar denotes the cost incurred during leaf node multiplication step and green bar signifies the cost incurred during divide and combine step. The cost signifies the dominant cost for each step i.e. computation cost for multiplication step and communication cost for divide and combine step.}\label{fig:Stagewise}
\end{figure*}

In this experiment, we compare the running time of three systems stage-wise. From this test, we can infer the time-consuming stage of each approach. We perform this test for increasing matrix size and for each matrix size with increasing partition size. As the number of stages of \textit{Stark} grows on the order of $(p-q)$, when there is a large difference between matrix size and partition size, the value of $p-q$ becomes too large to compare with other approaches. For this reason, we merge the stages of \textit{Stark} to form $3$ stages, comprising stages related to divide, leaf node multiplication and combine phases. Fig. \ref{fig:Stagewise} shows the comparison. To explain further, we have tabulated the stage-wise running time of the approaches with increasing partition size for all matrix sizes in Table \ref{tab:Stagewise-4096}, Table \ref{tab:Stagewise-8192}, and Table \ref{tab:Stagewise-16384}. The communication intensive stages are colored as green while computing intensive stages are filled with red.

It is clearly seen that, \textit{Stage 3} is the costliest stage for \textit{MLLib} and \textit{Marlin}. \textit{Stage 3} in \textit{MLLib} consists of \textit{coGroup} and \textit{flatMap} transformations, while in \textit{Marlin} it is \textit{partitionBy}, \textit{join} and \textit{mapPartitions} transformations. These contribute to the replication, shuffling and multiplications of blocks resulting cost far more than \textit{Stage4} having only one transformation \textit{reduceByKey}.

\begin{table*}[!ht]
\centering
	\caption{Stagewise performance comparison among three systems with increasing partition (The unit of execution time is second) for matrix size $4096\times 4096$}
	\label{tab:Stagewise-4096}
		\begin{center}
			\begin{tabular}{|c|c|c|c|c|c|c|c|c|c|c|c|c|}
				\hline
				\multirow{3}{*}{Stage} & \multicolumn{12}{c|}{Wall Clock Time} \\ \cline{2-13}
				& \multicolumn{3}{c|}{b=2} & \multicolumn{3}{c|}{b=4} & \multicolumn{3}{c|}{b=8} & \multicolumn{3}{c|}{b=16} \\ \cline{2-13}
				& \tiny MLLib & \tiny Marlin & \tiny Stark & \tiny MLLib & \tiny Marlin & \tiny Stark & \tiny MLLib & \tiny Marlin & \tiny Stark & \tiny MLLib & \tiny Marlin & \tiny Stark \\
				\hline			
				\multirow{2}{*}{Stage 1} & 4 & 1 & \multirow{2}{*}{8} & 5 & 1 & \multirow{2}{*}{4} & 1 & 1 & \multirow{2}{*}{4} & 2 & 2 & \multirow{2}{*}{4} \\
				& 4 & 1 & & 5 & 1 & & 2 & 0.9 & & 2 & 2 & \\
				\hline	
				\multirow{4}{*}{Stage 2} & - & - & \cellcolor{green}2 & - & - & \cellcolor{green}1 & - & - & \cellcolor{green}1 & - & - & \cellcolor{green}1 \\
				& - & - & - & - & - & \cellcolor{green}0.5 & - & - & \cellcolor{green}0.7 & - & - & \cellcolor{green}1 \\
				& - & - & - & - & - & - & - & - & \cellcolor{green}0.7 & - & - & \cellcolor{green}2 \\
				& - & - & - & - & - & - & - & - & - & - & - & \cellcolor{green}3 \\
				\hline	
				Stage 3 & \cellcolor{red}28 & \cellcolor{red}13 & \cellcolor{red}12 & \cellcolor{red}9 & \cellcolor{red}6 & \cellcolor{red}3 & \cellcolor{red}7 & \cellcolor{red}7 & \cellcolor{red}3 & \cellcolor{red}6 & \cellcolor{red}18 & \cellcolor{red}4 \\
				\hline	
				\multirow{4}{*}{Stage 4} & \cellcolor{green}0.2 & \cellcolor{green}1 & \cellcolor{green}0.9 & \cellcolor{green}0.1 & \cellcolor{green}0.8 & \cellcolor{green}0.4 & \cellcolor{green}0.2 & \cellcolor{green}2 & \cellcolor{green}0.3 & \cellcolor{green}0.1 & \cellcolor{green}10 & \cellcolor{green}0.6 \\
				& - & - & - & - & - & \cellcolor{green}0.6 & - & - & \cellcolor{green}0.3 & - & - & \cellcolor{green}0.3 \\
				& - & - & - & - & - & - & - & - & \cellcolor{green}0.4 & - & - & \cellcolor{green}0.2 \\
				& - & - & - & - & - & - & - & - & - & - & - &\cellcolor{green}0.3 \\		
				\hline	
			\end{tabular}
		\end{center}
\end{table*}

\begin{table*}[!ht]
	\caption{Stagewise performance comparison among three systems with increasing partition (The unit of execution time is second) for matrix size $8192\times 8192$}
	\label{tab:Stagewise-8192}
		\begin{center}
			\begin{tabular}{|c|c|c|c|c|c|c|c|c|c|c|c|c|}
				\hline
				\multirow{3}{*}{Stage} & \multicolumn{12}{c|}{Wall Clock Time} \\ \cline{2-13}
				& \multicolumn{3}{c|}{b=2} & \multicolumn{3}{c|}{b=4} & \multicolumn{3}{c|}{b=8} & \multicolumn{3}{c|}{b=16} \\ \cline{2-13}
				& \tiny MLLib & \tiny Marlin & \tiny Stark & \tiny MLLib & \tiny Marlin & \tiny Stark & \tiny MLLib & \tiny Marlin & \tiny Stark & \tiny MLLib & \tiny Marlin & \tiny Stark \\	
				\hline			
				\multirow{2}{*}{Stage 1} & 108 & 3 & \multirow{2}{*}{29} & 25 & 2 & \multirow{2}{*}{15} & 21 & 3 & \multirow{2}{*}{12} & 16 & 5 & \multirow{2}{*}{12} \\
				& 96 & 3 & & 29 & 2 & & 23 & 2 & - & 15 & 3 & \\
				\hline
				\multirow{4}{*}{Stage 2} & - & - & \cellcolor{green}4 & - & - & \cellcolor{green}4 & - & - & \cellcolor{green}4 & - & - & \cellcolor{green}4 \\
				& - & - & - & - & - & \cellcolor{green}2 & - & - & \cellcolor{green}3 & & & \cellcolor{green}3 \\
				& - & - & - & - & - & - & - & - & \cellcolor{green}3 & - & - & \cellcolor{green}5 \\
				& - & - & - & - & - & - & - & - & - & - & - & \cellcolor{green}5 \\
				\hline
				Stage 3 & \cellcolor{red}222 & \cellcolor{red}96 & \cellcolor{red}90 & \cellcolor{red}84 & \cellcolor{red}78 & \cellcolor{red}36 & \cellcolor{red}66 & \cellcolor{red}51 & \cellcolor{red}19 & \cellcolor{red}40 & \cellcolor{red}48 & \cellcolor{red}11 \\
				\hline
				\multirow{4}{*}{Stage 4} & \cellcolor{green}1 & \cellcolor{green}5 & \cellcolor{green}3 & \cellcolor{green}0.4 & \cellcolor{green}3 & \cellcolor{green}1 & \cellcolor{green}0.5 & \cellcolor{green}5 & \cellcolor{green}1 & \cellcolor{green}0.4 & \cellcolor{green}15 & \cellcolor{green}1 \\
				& - & - & - & - & - & \cellcolor{green}3 & - & - & \cellcolor{green}1 & - & - & \cellcolor{green}0.8 \\
				& - & - & - & - & - & - & - & - & \cellcolor{green}1 & - & - & \cellcolor{green}0.7 \\
				& - & - & - & - & - & - & - & - & - & - & - &\cellcolor{green}1 \\		
				\hline
			\end{tabular}
		\end{center}
\end{table*}

\begin{table*}[!ht]
	\caption{Stagewise performance comparison among three systems with increasing partition (The unit of execution time is second) for matrix size $16384\times 16384$}
	\label{tab:Stagewise-16384}
		\begin{center}
			\begin{tabular}{|c|c|c|c|c|c|c|c|c|c|}
				\hline
				\multirow{3}{*}{Stage} & \multicolumn{9}{c|}{Wall Clock Time} \\ \cline{2-10}
				& \multicolumn{3}{c|}{b=4} & \multicolumn{3}{c|}{b=8} & \multicolumn{3}{c|}{b=16} \\ \cline{2-10}
				& \tiny MLLib & \tiny Marlin & \tiny Stark & \tiny MLLib & \tiny Marlin & \tiny Stark & \tiny MLLib & \tiny Marlin & \tiny Stark \\	
				\hline			
				\multirow{2}{*}{Stage 1} & 150 & 9 & \multirow{2}{*}{90} & 49 & 14 & \multirow{2}{*}{48} & 33 & 43 & \multirow{2}{*}{43} \\
				& 156 & 6 & & 42 & 12 & & 34 & 48 &\\
				\hline
				\multirow{4}{*}{Stage 2} & - & - & \cellcolor{green}29 & - & - & \cellcolor{green}27 & - & - & \cellcolor{green}20 \\
				& - & - & \cellcolor{green}9 & - & - & \cellcolor{green}10 & - & - & \cellcolor{green}14 \\
				& - & - & - & - & - & \cellcolor{green}9 & - & - & \cellcolor{green}17 \\
				& - & - & - & - & - & - & - & - & \cellcolor{green}18 \\
				\hline
				Stage 3 & \cellcolor{red}720 & \cellcolor{red}582 & \cellcolor{red}384 & \cellcolor{red}516 & \cellcolor{red}504 & \cellcolor{red}192 & \cellcolor{red}474 & \cellcolor{red}348 & \cellcolor{red}78 \\
				\hline
				\multirow{4}{*}{Stage 4} & \cellcolor{green}2 & \cellcolor{green}23 & \cellcolor{green}8 & \cellcolor{green}2 & \cellcolor{green}17 & \cellcolor{green}3 & \cellcolor{green}1 & \cellcolor{green}34 & \cellcolor{green}4 \\
				& - & - & \cellcolor{green}11 & - & - & \cellcolor{green}7 & - & - & \cellcolor{green}2 \\
				& - & - & - & - & - & \cellcolor{green}5 & - & - & \cellcolor{green}3 \\
				& - & - & - & - & - & - & - & - &\cellcolor{green}5 \\		
				\hline
			\end{tabular}
		\end{center}
\end{table*}

For \textit{Stark}, on the other hand, the most expensive stage changes as we increase partition size. For smaller partition size \textit{Stage 3} or leaf node multiplication computation cost dominates while for larger partitions \textit{Stage 2} or communication cost of matrix division dominates. This is because, the communication cost corresponding to the divide section dominates as $b$ is increased and as the height of the recursion tree increases, the communication cost accumulates and surpasses the computation cost of leaf node multiplications. The main factor that makes \textit{Stark} to be faster is its ability to preserve the number of multiplications to be smaller than the other two approaches. It is clear from tables \ref{tab:Stagewise-4096}, \ref{tab:Stagewise-8192} and \ref{tab:Stagewise-16384}, the most costly stage is the stage that contains the block matrix multiplication calculation. This cost is almost same for smaller partition size, but the gap increases as we move to larger partition size. The reason is that, $(b^{2.8}\times (\frac{n}{b})^{3})/min[b^{2.8},cores])$ factor in \textit{Stark} grows slowly than the factor $(b^{3}\times (n/b)^{3}/min[b^{3},cores])$ in \textit{Marlin}. As we increase the partition size $b$, the number of multiplications for \textit{Marlin} to be carried-out grows in $b^{3}$ order, while \textit{Stark} grows in $7^{p-q}$ or $b^{2.8}$ order, which is less than the earlier. Again, \textit{Stark} makes the division and combination phase to be parallel enough, making it faster compared to any other approaches. 

\begin{figure*}[!ht]
\centering
	\begin{subfigure}[t]{.3\textwidth}
		\begin{tikzpicture}
		\begin{axis}[
		xmin=0,
		xmax=6,
		xtick={1,2,3,4,5},
		ytick={4,8,12,16,20,24},
		x tick label style={/pgf/number format/1000 sep=},
		xlabel={\tiny Number of Executors},
		ylabel={\tiny Wall Clock Time in Sec.},
		legend style={at={(0.5,1.5)},anchor=north},
		width=\textwidth,
		]
		\addplot coordinates {(1,21.5) (2,10.75) (3,7.16) (4,5.37) (5,4.3)};
		\addlegendentry{Ideal}
		\addplot coordinates {(1,21.5) (2,14) (3,10.8) (4,10.8) (5,9.5)};
		\addlegendentry{Experimental}
		\end{axis}
		\end{tikzpicture}
		\caption{$4096\times 4096$},
	\end{subfigure}%
	\begin{subfigure}[t]{.3\textwidth}
		\begin{tikzpicture}
		\begin{axis}[
		xmin=0,
		xmax=6,
		xtick={1,2,3,4,5},
		ytick={25,50,75,100,125,150,175,200},
		x tick label style={/pgf/number format/1000 sep=},
		xlabel={\tiny Number of Executors},
		ylabel={\tiny Wall Clock Time in Sec.},
		legend style={at={(0.5,1.5)},anchor=north},
		width=\textwidth,
		]
		\addplot coordinates {(1,114.6) (2,57.3) (3,38.2) (4,28.65) (5,22.92)};
		\addlegendentry{Ideal}
		\addplot coordinates {(1,114.6) (2,66.6) (3,50.7) (4,46.4) (5,40.6)};
		\addlegendentry{Experimental}
		\end{axis}
		\end{tikzpicture}
		\caption{$8192\times 8192$}
	\end{subfigure}%
	\begin{subfigure}[t]{.3\textwidth}
		\begin{tikzpicture}
		\begin{axis}[
		xmin=0,
		xmax=6,
		xtick={1,2,3,4,5},
		ytick={250,500,750,1000,1250,1500},
		x tick label style={/pgf/number format/1000 sep=},
		xlabel={\tiny Number of Executors},
		ylabel={\tiny Wall Clock Time in Sec.},
		legend style={at={(0.5,1.5)},anchor=north},
		width=\textwidth,
		]
		\addplot coordinates {(1,1525) (2,762.5) (3,508.33) (4,381.25) (5,305)};
		\addlegendentry{Ideal}
		\addplot coordinates {(1,1525) (2,763) (3,520) (4,410) (5,325)};
		\addlegendentry{Experimental}
		\end{axis}
		\end{tikzpicture}
		\caption{$16384\times 16384$}
	\end{subfigure}%
	\caption{The scalability of \textit{Stark}, in comparison with ideal scalability (blue line), on matrix $(4096\times 4096)$, $(8192\times 8192)$ and $(16384\times 16384)$. The wall clock execution time is recorded against increasing number of executors (the number of cores is $5$ for all the executors i.e. $1-5$. Here ideal scalability is $T(1)/n$, where $T(1)$ is wall clock execution time when number of executors $=1$ and $n$ is total number of executors.}\label{fig:scale}
\end{figure*}
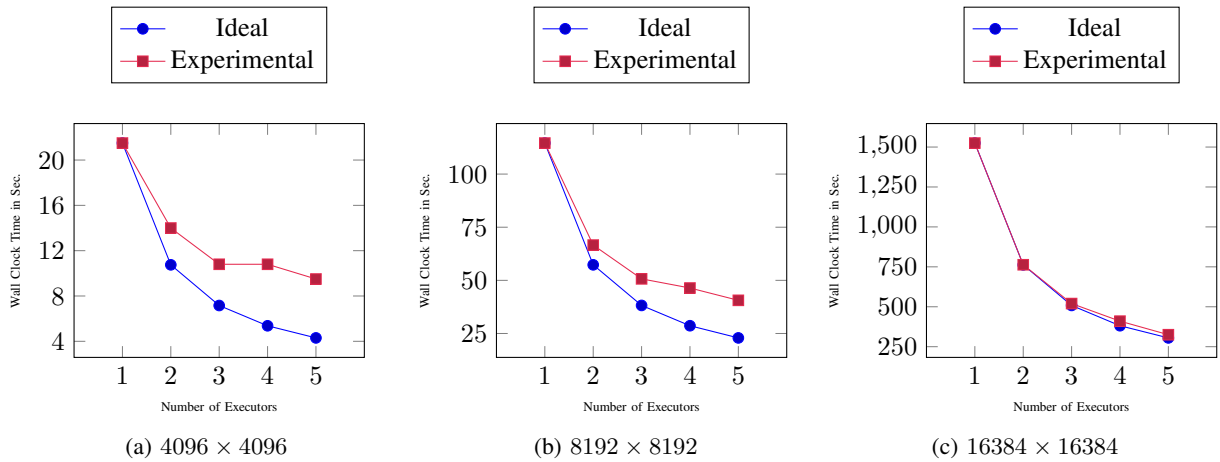

\subsection{Scalability}
In this section, we investigate the scalability of \textit{Stark}. For this, we generate three test cases, each containing a different set of two matrices of dimensions equal to $(4096\times 4096)$, $(8192\times 8192)$ and $(16384\times 16384)$. The running time vs. the number of spark executors for these $3$ pairs of matrices is shown in Fig. \ref{fig:scale}. The ideal scalability line (i.e. $T(n) = T(1)/n$ - where $n$ is the number of executors) has been over-plotted on this figure in order to demonstrate the scalability of our algorithm. We can see that \textit{Stark} has a strong scalability, with a minor deviation from ideal scalability when the size of the matrix is low (i.e. for $(8192\times 8192)$ and $(16384\times 16384)$).  

\section{Conclusions and Future Work}
\label{sec:conclusion}
In this paper, we have focused on the problem of distributed matrix multiplication of large and distributed matrices using Spark framework. Here, we have overcome the shortcomings in the state-of-the-art distributed matrix multiplication approaches requiring $O(n^{3})$ running time. We have accomplished that by providing an efficient distributed implementation of the sub-cubic $O(n^{2.807})$ time, Strassen's multiplication algorithm. A key novelty is to simulate the distributed recursion by carefully tagging the matrix blocks and processing each level of the recursion tree in parallel.
We have also report a comprehensive theoretical analysis of the computation and communication costs and parallization factor associated with each stage of \textit{Stark} as well as other baseline approaches. Through extensive experiments on the wall clock execution time of the competitive approaches and find that the theoretical analysis matches with the empirical one and also pinpoint the actual source of improvement in wall clock time of \textit{Stark}.

An important drawback of the current approach is it's high space complexity, which is $O(3^l N^2)$, where $l$ is the recursion level (this is because at each recursion level the size of the data grows $3$ times than the previous level). For matrix having a million rows and columns will experience a huge memory consumption and thus cannot be executed in situations when other approaches can be run smoothly. It will be interesting to find way to circumvent this problem. 


%





\ifCLASSOPTIONcaptionsoff
  \newpage
\fi



\bibliographystyle{IEEEtran}
\bibliography{bibtex/bib/IEEEexample.bib}

%

\begin{IEEEbiography}[{\includegraphics[width=1in,height=1.5in,clip,keepaspectratio]{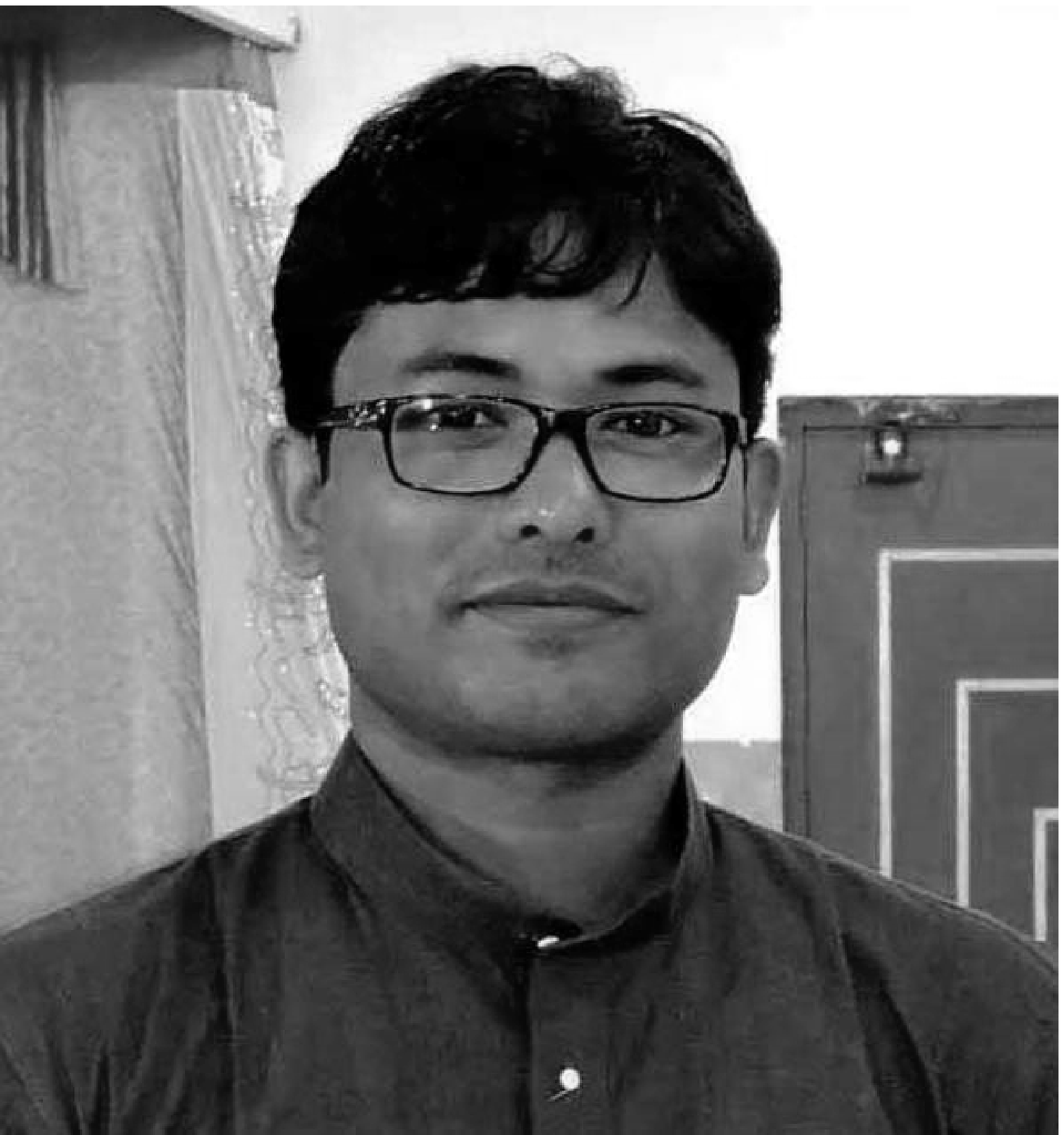}}]{Chandan Misra}
received the B.Tech degree in information technology from West Bengal University of Technology, West Bengal, India, and the M.S degree in Computer Science from the Indian Institute of Technology (IIT) Kharagpur, India. He is currently working toward the Ph.D. degree from the
Advanced Technology Development Centre, IIT Kharagpur, Kharagpur, India. 

His research interests include spatial data mining and information retieval, big data, and distributed linear algebra.
\end{IEEEbiography}

\begin{IEEEbiography}[{\includegraphics[width=1in,height=1.25in,clip,keepaspectratio]{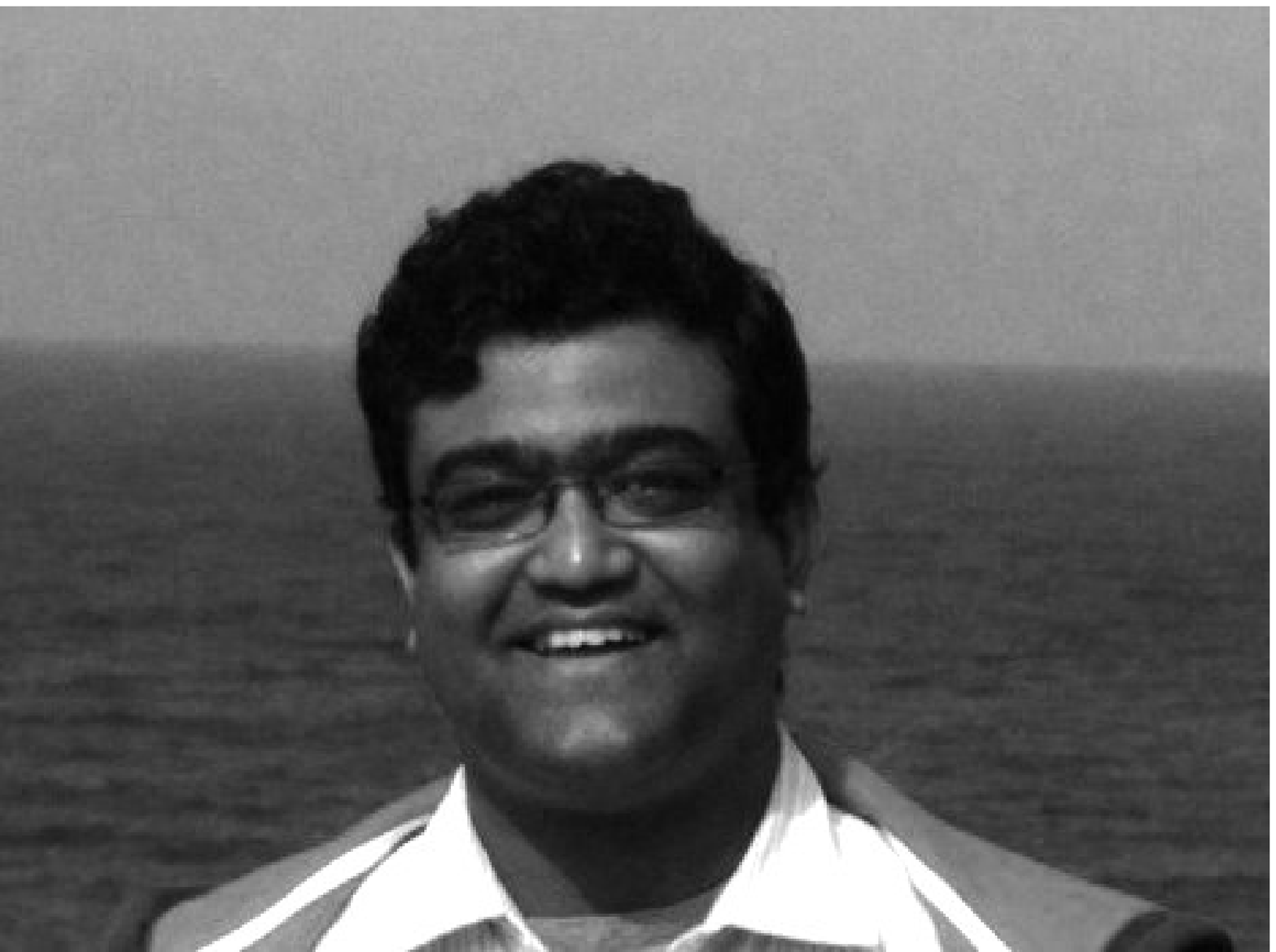}}]{Sourangshu Bhattacharya} is an Assistant Professor in the Department of Computer Science and Engineering, IIT Kharagpur. He was a Scientist at Yahoo! Labs from 2008 to 2013, where he was working on prediction of Click-through rates, Ad-targeting to customers, etc on the Rightmedia display ads exchange. He was a visiting scholar at the Helsinki University of Technology from January - May 2008. He received the B.Tech. in Civil Engineering from I.I.T. Roorkee in 2001, M.Tech. in computer science from I.S.I. Kolkata in 2003, and Ph.D. in Computer Science from the Department of Computer Science \& Automation, IISc Bangalore in 2008. He has many publications in top conferences and journals. His current research interests include applications of machine learning in Opinion Dynamics in Online Social networks, Online advertising, Information Extraction and Transportation Science; as well as in distributed machine learning, and representation learning.
\end{IEEEbiography}

\begin{IEEEbiography}[{\includegraphics[width=1in,height=1.25in,clip,keepaspectratio]{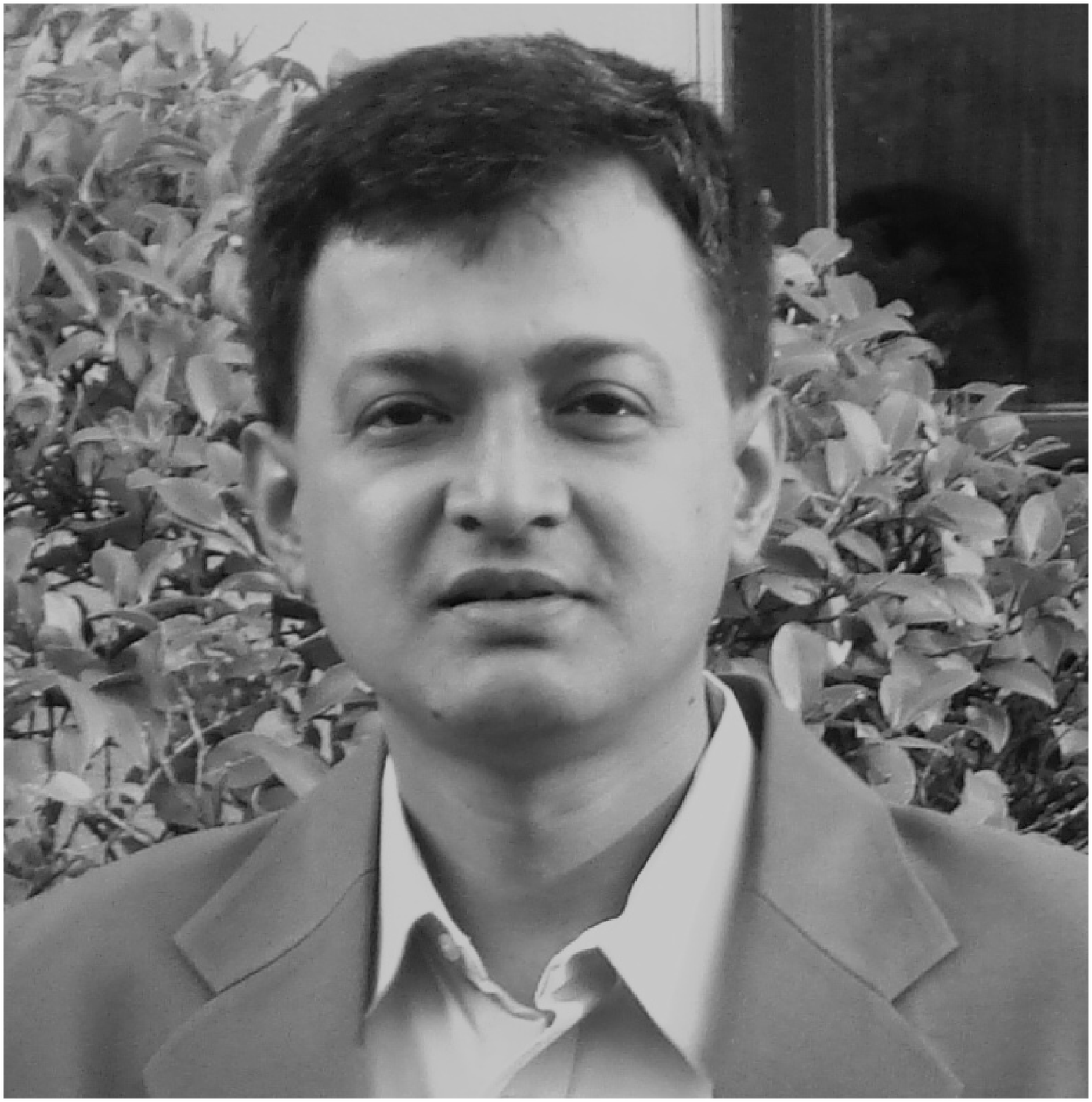}}]{Soumya K. Ghosh}
Soumya K. Ghosh received the M.Tech. and Ph.D. degrees in computer science and engineering from the Indian Institute of Technology (IIT) Kharagpur, Kharagpur, India. He is currently a Professor with the Department of Computer Science and Engineering, IIT Kharagpur. Before joining IIT Kharagpur, he was with the Indian Space Research Organization, working in the area of satellite remote sensing and geographic information system (GIS). He has over $100$ research publications in journals and conference proceedings. His research interests include geoscience and spatial web services, mobile GIS, spatial information retrieval, and knowledge discovery.
\end{IEEEbiography}






\enlargethispage{-5in}

\end{document}